\documentstyle[prd,epsfig,preprint,aps,tighten]{revtex}
\newcommand{\LT}        {\left}
\newcommand{\RT}        {\right}
\newcommand{\0}         {\hphantom{0}}

\renewcommand{\ne}      {\hbox{$\nu_e$} }
\newcommand{\nm}        {\hbox{$\nu_\mu$} }

\def \lsim {\mbox{${}^< \hspace*{-7pt} _\sim$}}
\def \gsim {\mbox{${}^> \hspace*{-7pt} _\sim$}}
\begin{document}
\begin{titlepage}
\preprint{
         \vbox{\hbox{hep-ph/0009350}
               \hbox{IFIC/00-51}}}
\title{Global three--neutrino oscillation analysis of  neutrino data}
\author{
  M.~C.~Gonzalez-Garcia\thanks{E-mail: \tt concha@flamenco.ific.uv.es}, 
  M.~Maltoni\thanks{E-mail: \tt maltoni@flamenco.ific.uv.es},
  C.~Pe\~na-Garay\thanks{E-mail: \tt penya@flamenco.ific.uv.es},
  and J.~W.~F.~Valle\thanks{E-mail: \tt valle@flamenco.ific.uv.es}
  }
\address{
  Instituto de F\'{\i}sica Corpuscular~--~C.S.I.C., Universitat de Val\`encia, \\
  Edificio Institutos de Paterna, Apt.~22085, E-46071 Val\`encia, Spain}
\maketitle
\vskip -0.75cm
\begin{abstract}
\vskip -.75cm
\baselineskip 0.425 cm
  A global analysis of the solar, atmospheric and reactor neutrino
  data is presented in terms of three--neutrino oscillations. We
  include the most recent solar neutrino rates of Homestake, SAGE,
  GALLEX and GNO, as well as the recent 1117 day Super--Kamiokande data
  sample, including the recoil electron energy spectrum both for day
  and night periods and we treat in a unified way the full parameter
  space for oscillations, correctly accounting for the transition from
  the matter enhanced (MSW) to the vacuum oscillations regime.
  Likewise, we include in our description conversions with
  $\theta_{12} > \pi/4$.
  For the atmospheric data we perform our analysis of the contained
  events and the upward-going $\nu$-induced muon fluxes,
  including the previous data samples of Frejus, IMB, Nusex, and
  Kamioka experiments as well as the full 71 kton-yr (1144 days)
  Super--Kamiokande data set, the recent 5.1 kton-yr
  contained events of Soudan2 and the results on upgoing muons from
  the MACRO detector.
  We first present the allowed regions of solar and atmospheric
  oscillation parameters $\theta_{12}$, $\Delta m^2_{21}$ and
  $\theta_{23}$, $\Delta m^2_{32}$, respectively, as a function of
  $\theta_{13}$.  We determine the constraints from atmospheric and
  solar data on the mixing angle $\theta_{13}$, common to solar and
  atmospheric analyses. The solar limit on $\theta_{13}$, although
  relatively weak, is totally independent on the allowed range of the
  atmospheric mass difference $\Delta m^2_{32}$. On the other hand the
  atmospheric data analysis indicates an important complementarity
  with the reactor limits allowing for a stronger constraint on the
  allowed value of $\theta_{13}$.  We also obtain the allowed ranges
  of parameters from the full five--dimensional combined analysis of
  the solar, atmospheric and reactor data.
\end{abstract}
\end{titlepage}
\section{Introduction}

Super--Kamiokande high statistics data~\cite{skcont,skup,skatm00}
indicate that the observed deficit in the $\mu$-like atmospheric
events is due to the neutrinos arriving in the detector at large
zenith angles, strongly suggestive of the \nm oscillation hypothesis.
Similarly, their data on the zenith angle dependence and recoil energy
spectrum of solar neutrinos~\cite{sksol,sksol00} in combination with
the results from Homestake~\cite{chlorine}, SAGE~\cite{sage}, and
GALLEX+GNO~\cite{gallex,gno} experiments, have put on a firm
observational basis the long--standing problem of solar neutrinos,
strongly indicating the need for \ne~conversions.

Altogether, the solar and atmospheric neutrino
anomalies~\cite{skcont,skup,skatm00,sksol,sksol00,IMB,Kamiokande,Soudan2,MACRO}
constitute the only solid present--day evidence for physics beyond the
Standard Model (SM). 
It is clear that the minimum
joint description of both anomalies requires neutrino conversions
amongst all three known neutrinos. In the simplest case of
oscillations the latter are determined by the structure of the lepton
mixing matrix~\cite{MNS}, which,
in addition to the Dirac-type phase
analogous to that of the quark sector contains two
physical~\cite{Schechter:1981gk} phases associated to the Majorana character of
neutrinos.  
CP conservation implies that lepton phases are either zero
or $\pi$~\cite{Schechter:1981hw}. For our following description it
will be correct and sufficient to set all three phases to zero.  In
this case the mixing matrix can be conveniently chosen in the 
form~\cite{PDG}
\begin{equation}
{\bf R} =\left(
    \begin{array}{ccc}                                 
        c_{13} c_{12}                
        & s_{12} c_{13} 
        & s_{13} \\
        -s_{12} c_{23} - s_{23} s_{13} c_{12} 
        & c_{23} c_{12} - s_{23} s_{13} s_{12}
        & s_{23} c_{13} \\
        s_{23} s_{12} - s_{13} c_{23} c_{12}
        & -s_{23} c_{12} - s_{13} s_{12} c_{23}
        & c_{23} c_{13}
    \end{array}\right) \;,
\label{eq:evol.2} 
\end{equation}
With this the parameter set relevant for the joint study of solar and
atmospheric conversions becomes five-dimensional:
\begin{equation} \begin{array}{ll} 
   \label{oscpardef}
    \Delta m^2_{\odot} & \equiv \Delta m^2_{21} = m^2_2 - m^2_1 \\
    \Delta m^2_{atm}   & \equiv \Delta m^2_{32} = m^2_3 - m^2_2 \\
    \theta_{\odot}     & \equiv \theta_{12} \\
    \theta_{atm}       & \equiv \theta_{23} \\
    \theta_{reactor}   & \equiv \theta_{13}
\end{array} 
\end{equation}
where all mixing angles are assumed to lie in the full range from
$[0,\pi/2]$. 

In this paper we present a global analysis of the data on solar,
atmospheric and reactor neutrinos, in terms of three--family neutrino
oscillations. There are several three--neutrino oscillation analyses
in the literature which either include solar~\cite{3solmsw,3solvac} or
atmospheric neutrino data~\cite{3atm}. Joint studies were also
performed, but without including the most recent and precise
Super--Kamiokande data~\cite{3old}.  This work updates and combines
all these results in a unique comprehensive analysis.  It is known
that in the case  $\Delta m^2_{32} \gg \Delta m^2_{12}$,  
for $\theta_{13}=0$ the atmospheric and solar neutrino
oscillations decouple in two two--neutrino oscillation scenarios. In
this respect our results also contain as limiting cases the pure
two--neutrino oscillation scenarios and update previous analyses on
atmospheric neutrinos~\cite{2atmour,2atmothers} and solar
neutrinos~\cite{2solour,2solother} (for an updated analysis of on
two--neutrino oscillation of solar neutrino data see also Ref.~\cite{nu2000}).

We include in our analysis the most recent solar neutrino rates of
Homestake~\cite{chlorine}, SAGE~\cite{sage}, GALLEX and
GNO~\cite{gallex,gno}, as well as the recent 1117 day
Super--Kamiokande data sample~\cite{sksol00}, including the recoil
electron energy spectrum both for day and night periods. Treating in a
unified way the full parameter space for oscillations, we carefully
account for the transition from the matter enhanced to the vacuum
oscillations regime.  Likewise, we include MSW conversions in the
second octant ({\it dark-side}), characterized by $\theta_{12} >
\pi/4$~\cite{3solmsw,3solvac,dark}.
As for atmospheric neutrinos we include in our analysis all the
contained events as well as the upward-going neutrino-induced muon
fluxes, including both the previous data samples of
Frejus~\cite{Frejus}, IMB~\cite{IMB}, Nusex~\cite{Nusex} and Kamioka
experiments ~\cite{Kamiokande} and the full 71 kton-yr
Super--Kamiokande data set~\cite{skatm00}, the recent 5.1 kton-yr
contained events of Soudan2~\cite{Soudan2} and the results on upgoing
muons from the MACRO detector~\cite{MACRO}.  We also determine the
constraints implied by the CHOOZ reactor experiments ~\cite{CHOOZ}.

From the required hierarchy in the splittings 
$\Delta m^2_{atm} \gg \Delta m^2_{\odot}$ 
indicated by the solutions to the solar and
atmospheric neutrino anomalies (which indeed we justify {\it a
  posteriori}) it follows that the analyses of solar data constrain
three of the five independent oscillation parameters, namely, $\Delta
m^2_{21}, \theta_{12}$ and $\theta_{13}$. Conversely, the atmospheric
data analysis restricts $\Delta m^2_{32}$, $\theta_{23}$ and
$\theta_{13}$, the latter being the only parameter common to both and
which may potentially allow for some mutual influence. In our global
approach we will statistically combine these solar and atmospheric
limits on $\theta_{13}$. We also compare this bound and combine them
with the direct limit on $\theta_{13}$ which follows from reactor
experiments.  While the solar limits on $\theta_{13}$ are relatively
weak, our atmospheric data analysis indicates an important
complementarity between the atmospheric and the reactor limits.

The outline of the paper is the following: in Sec.~\ref{probs} we
review the theoretical calculation of the conversion probabilities for
solar, atmospheric and reactor neutrinos in the framework of
three--neutrino mixing; in Sec.~\ref{data} we describe the data
samples, the computation of the theoretical observables, and the
statistical analysis applied in our analysis for solar
(Sec.~\ref{data:solar}), atmospheric (Sec.~\ref{data:atm}), and
reactor~\ref{data:CHOOZ} data. Section~\ref{analysis:solar} is devoted
to our results for the three--neutrino oscillation fits to solar
neutrino data. Correspondingly in Sec.~\ref{analysis:atmos} we
describe our results for atmospheric neutrino fits by themselves and
also in combination with the reactor data.  The results for the full
combined five--parameter analysis are described in
Sec.~\ref{analysis:combined}. Finally in Sec.~\ref{sum} we summarize
the work and present our conclusions.

\section{Three--neutrino Oscillation Probabilities}
\label{probs}

In this section we review the theoretical calculation of the
conversion probabilities for solar, atmospheric and reactor neutrinos
in the framework of three--neutrino mixing in order to set our
notation and to clarify the approximations used in the evaluation of
such probabilities.

In general, the determination of the oscillation probabilities both
for solar neutrinos and atmospheric neutrinos require the solution of
the Schr\"oedinger evolution equation of the neutrino system in the
Sun-- and/or the Earth--matter background. For a three-flavour
scenario, this equation reads
\begin{equation}
    i \frac{d\vec{\nu}}{dt} = {\bf H} \, \vec{\nu},\;\;\;\;\;\;\;\; 
     {\bf H} = {\bf R} \cdot {\bf H}_0^d \cdot {\bf R}^\dagger 
    + {\bf V} \;,
\label{eq:evol.1} 
\end{equation}
where ${\bf R}$ is the orthogonal matrix connecting the flavour basis
and the mass basis in vacuum and which can be parametrized as in
Eq.~(\ref{eq:evol.2}). On the other hand ${\bf H}_0^d$ and  {\bf V}
are given as
\begin{equation}
    {\bf H}_0^d = \frac{1}{2 E_\nu} {\bf diag}
    \LT( - \Delta m^2_{21}, 0, \Delta m^2_{32} \RT),
\label{eq:evol.3} 
\end{equation}
\begin{equation}
    {\bf V} = 
    {\bf diag} \LT( \pm \sqrt{2} G_F N_e, 0, 0 \RT),
 \label{eq:evol.4} 
\end{equation}
where $\vec{\nu} \equiv \LT( \nu_e, \nu_\mu, \nu_\tau \RT)$, $c_{ij}
\equiv \cos\theta_{ij}$ and $s_{ij} \equiv \sin\theta_{ij}$.
The angles $\theta_{ij}$ can be taken without any loss of generality
to lie in the first quadrant $\theta_{ij}\in[0,\pi/2]$.  We have
denoted by ${\bf H}_0^d$ the vacuum hamiltonian, while ${\bf V}$
describes charged-current forward interactions in matter.
In Eq.~(\ref{eq:evol.4}), the sign $+$ ($-$) refers to neutrinos
(antineutrinos), $G_F$ is the Fermi coupling constant and $N_e$ is
electron number density in the Sun or the Earth.

In writing Eq.~(\ref{eq:evol.2}) we have set all three CP violating
phases to zero. Although this is, in general, an approximation, it
holds exact for the scheme we are adopting in the description of 
solar and atmospheric data
\begin{equation} 
\Delta m^2_{21}\ll 
\Delta m^2_{32}\simeq\Delta m^2_{31} \;,
\label{hierarchical}
\end{equation}
because in this case, as we describe below, no simultaneous effect of
the two mass differences is observable in any $\nu$--appearance
transition. This is the case, for instance, for the hierarchical scheme,
$ m_1<m_2\ll m_3$.
Notice also that for transitions in vacuum the results
obtained apply also to the inverted hierarchical case
$m_1>m_2\gg m_3 $.
In the presence of matter effects the hierarchical and inverted
hierarchical cases are no longer equivalent, although as discussed in
Ref.~\cite{3atm} the difference is hardly recognizable in the current
solar and atmospheric neutrino phenomenology.

\subsection{Solar Neutrinos}

For solar neutrinos we can write the survival amplitude for $\nu_e$
neutrinos of energy $E$ at a detector in the Earth as
\begin{equation}
A_{ee}=\sum_{i=1}^3 A^S_{e\,i}\,A^E_{i\,e}\,\exp[-im_i^2 (L-r)/2E]~. 
\end{equation}
Here $A^S_{e\,i}$ is the amplitude of the $\nu_e \to \nu_i$ transition
($\nu_i$ is the $i$-mass eigenstate) from the production point to the
Sun surface, $A^E_{i\,e}$ is the amplitude of the transition $\nu_i
\to \nu_e$ from the Earth surface to the detector, and the propagation
in vacuum from the Sun to the surface of the Earth is described by the
exponential.  $L$ is the distance between the center of the Sun and
the surface of the Earth, and $r$ is the distance between the neutrino
production point and the surface of the Sun.  Using the mass hierarchy
in Eq.~(\ref{hierarchical}) (which also implies that for the evolution
both in the Sun and in the Earth $\Delta m^2_{32}\gg
2\sqrt{2} G_F N_e E_\nu\sin^2(2\theta_{13})$ the three--flavour evolution 
equations decouple into an effective two--flavour problem for the intermediate
basis~\cite{kuo-panta,shi-schramm}
\begin{equation}
\nu_{e'} = \cos \theta_{12} ~\nu_1 + \sin \theta_{12} ~\nu_2 \; , \;
 \nu_{\mu'} = - \sin \theta_{12} ~\nu_1 + \cos \theta_{12} ~\nu_2  \; ,
\label{eigendef}
\end{equation} 
with the substitution of $N_e$ by the effective density 
\begin{equation}
N_{e'}\Rightarrow N_e \cos^2 \theta_{13} \; ,
\label{pote}
\end{equation}
while for the evolution of the third state $\nu_{\tau'}= \nu_3$ there 
are no matter effects. 
Thus, the survival amplitude can be simplified to
\begin{equation}
A_{ee}=\cos^2 \theta_{13} \sum_{i=1}^2 A^S_{e'\,i}\,A^E_{i\,e'}\,
\exp[-im_i^2 (L-r)/2E] + \sin^2 \theta_{13}\,\exp[-im_3^2 (L-r)/2E] ~. 
\end{equation}
The survival probability after averaging out the interference terms 
due to the higher mass difference $\Delta m^2_{32}\simeq\Delta m^2_{31}$ 
is given by:
\begin{equation}
P^{3\nu}_{ee}=\cos^4 \theta_{13} P^{2\nu}_{e'e'}+\sin^4 \theta_{13} ~,
\label{Peesol}
\end{equation}
where we label $P^{2\nu}_{e'e'}$ the corresponding two--flavour survival
probability in the ($\Delta m^2_{21}, \theta_{12}$) parameter space 
but with the modified matter density in Eq.~(\ref{pote}). 
The expression for this effective two--flavour survival probability 
can be expressed as:
\begin{equation} 
P^{2\nu}_{e'e'}=P_1P_{1e'}+P_2P_{2e'}+2\sqrt{P_1P_2P_{1e'}P_{2e'}}\cos\xi \; .
\label{Pee}
\end{equation}
Here $P_i\equiv |A^S_{e'\,i}|^2$, while 
$P_{ie'} \equiv  |A^E_{i\,e'}|^2$ and unitarity implies that 
$P_1+P_2=1$ and $P_{1e'}+P_{2e'}=1$.  
The phase $\xi$ is given by 
\begin{equation} 
\xi=\frac{\Delta m^2_{21} (L-r)}{2E}+\delta\, ,
\end{equation}
where $\delta$ contains the phases due to propagation in the
Sun and in the Earth and can be safely neglected.
In the evaluation of both $P_1$ and $P_{2e'}$ the effect of coherent
forward interaction with the Sun and Earth matter is taken
into account with the effective density in Eq.~(\ref{pote}).

From  Eq.~(\ref{Pee}) one can recover more familiar expressions for 
$P^{2\nu}_{e'e'}$:  

(1) For $\Delta m^2_{21}/E\lsim 5 \times 10^{-17}$ eV, the matter
effect suppresses flavour transitions both in the Sun and the Earth.
Consequently, the probabilities $P_1$ and $P_{2e'}$ are simply the
projections of the $\nu_{e'}$ state onto the mass eigenstates: $P_1 =
\cos^2\theta_{12}$, $P_{2e'} = \sin^2 \theta_{12}$.  In this case we
are left with the standard vacuum oscillation formula:
\begin{equation}
P_{e'e'}^{2\nu,\rm vac}=1-\sin^2(2\theta _{12})\sin^2(\Delta m^2_{21} (L-r)/4E)
\label{pvac}
\end{equation}
which describes the oscillations on the way from the surface of the
Sun to the surface of the Earth.

(2) For $\Delta m^2_{21}/E\gtrsim 10^{-14}$ eV, the last term in
Eq.~(\ref{Pee}) vanishes and we recover the incoherent MSW survival
probability.  For $\Delta m^2_{21}/E\sim10^{-14}-10^{-12}$ eV$^2$,
this term is zero because $\nu_{e'}$ adiabatically converts to $\nu_2$
and $P_1=0$.  For $\Delta m^2_{21}/E\gsim10^{-12}$ eV$^2$, both $P_1$
and $P_2$ are nonzero and the term vanishes due to averaging of
$\cos\xi$.

(3) In the intermediate range, $5 \times 10^{-17} \lesssim \Delta
m^2_{21}/E \lesssim 10^{-14}$ eV, adiabaticity is violated and the
$\cos\xi$ coherent term should be taken into account. The result is
similar to vacuum oscillations but with small matter corrections. We
define this case as quasi-vacuum
oscillations~\cite{3solvac,nu2000,petcov1,panta,maxmix,petcov2,fried}.

In order to compute the survival probability for solar neutrinos,
valid for any value of the neutrino mass and mixing, the full
expression (\ref{Peesol}) has to be used. The results presented
in the following sections have been obtained using the general
expression for the survival probability in 
Eqs.~(\ref{Peesol})and~(\ref{Pee}) with $P_1$
and $P_{2e'}$ found by numerically solving the evolution equation in
the Sun and the Earth matter. For $P_1$ we use the electron number
density of BP2000 model~\cite{BP00}. For $P_{2e'}$ we integrate
numerically the evolution equation in the Earth matter using the Earth
density profile given in the Preliminary Reference Earth Model
(PREM)~\cite{PREM}.

\subsection{Atmospheric and Reactor Neutrinos}

For the atmospheric neutrino analysis it is a good
approximation to take two of the neutrinos as approximately
degenerate, given the hierarchy in the splittings $\Delta m^2_{atm}$
and $\Delta m^2_{\odot}$ which is indicated by the solutions to the
solar and atmospheric neutrino anomalies. In the $\Delta m^2_{21} \to
0$ approximation one can rotate away the corresponding angle
$\theta_{12}$, leading to the following expression for the leptonic
mixing matrix in vacuum~\cite{Schechter:1980bn}
\begin{equation}
    {\bf R} =\left(
    \begin{array}{ccc}                                 
        c_{13}          & 0       & s_{13} \\
        - s_{23} s_{13} & c_{23}  & s_{23} c_{13} \\
        - s_{13} c_{23} & -s_{23} & c_{23} c_{13}
    \end{array}\right); 
\label{eq:evolap.1}
\end{equation}
\begin{equation}
    {\bf H}_0^d = \frac{1}{2 E_\nu} {\bf diag} 
    \LT( 0, 0, \Delta m^2_{32} \RT).
\label{eq:evolap.2}
\end{equation}
As a result the 3-neutrino propagation of atmospheric neutrinos can be
well described by only three oscillation parameters: $\Delta m^2_{32}$,
$\theta_{23}$ and $\theta_{13}$.

For $\theta_{13}$=0, atmospheric neutrinos involve only $\nu_\mu \to
\nu_\tau$ conversions, and in this case there are no matter effects,
so that the solution of Eq.~(\ref{eq:evol.1}) is straightforward and
the conversion probability takes the well-known vacuum form
\begin{equation}
    P_{\mu\mu} = 1 - \sin^2 \LT( 2 \theta_{23} \RT)
    \sin^2 \LT( \frac{\Delta m^2_{32} L}{4 E_\nu} \RT),
\end{equation}
where $L$ is the path-length travelled by neutrinos of energy $E_\nu$.

On the other hand, in the general case of three-neutrino scenario with
$\theta_{13}\neq 0$ the presence of the matter potentials requires a
numerical solution of the evolution equations in order to obtain the
oscillation probabilities for atmospheric neutrinos $P_{\alpha\beta}$,
which are different for neutrinos and anti-neutrinos because of the
reversal of sign in Eq.~(\ref{eq:evol.4}). In our calculations, we
will use for the matter density profile of the Earth the approximate
analytic parametrization given in Ref.~\cite{lisi} of the PREM model
of the Earth~\cite{PREM}.

As for the CHOOZ reactor data, we need to evaluate the survival
probability for ${\bar \nu}_e$ of average energy $E\sim$ few MeV at a
distance of $L\sim 1$ Km. For this value of energy and distance one
can compute the survival probability neglecting Earth matter effects. 
In this case the survival probability takes the analytical form:
\begin{eqnarray}
P_{ee}^{CHOOZ}&=& 1-
\cos^4\theta_{13}
\sin^2(2\theta_{12})
\sin^2 \LT( \frac{\Delta m^2_{21} L}{4 E_\nu} \RT) \\ \nonumber
&  & -\sin^2(2\theta_{13})
\left(\cos^2\theta_{12}\sin^2 \LT( \frac{\Delta m^2_{31} L}{4 E_\nu} \RT)
+\sin^2\theta_{12}\sin^2 \LT( \frac{\Delta m^2_{32} L}{4 E_\nu} \RT) \right)
\\\nonumber
& \simeq &  
1-\sin^2(2\theta_{13})\sin^2 \LT( \frac{\Delta m^2_{32} L}{4 E_\nu} \RT)
,
\label{pchooz}
\end{eqnarray}
where the second equality holds under the  approximations in
Eqs.~(\ref{eq:evolap.1}) and~(\ref{eq:evolap.2}) and is fully valid 
for $\Delta m^2_{21}\lesssim 3\times 10^{-4}$eV$^2$.
\section{Data and Statistical Analysis}
\label{data}
\subsection{Solar Neutrinos}
\label{data:solar}

In order to determine the values of neutrino masses and mixing for the
oscillation solution of the solar neutrino problem, we have used data
on the total event rates measured in the Chlorine experiment at
Homestake~\cite{chlorine}, in the two Gallium experiments GALLEX+GNO
and SAGE~\cite{sage,gallex,gno} and in the water Cerenkov detectors
Kamiokande~\cite{kamioka} and Super--Kamiokande~\cite{sksol00} shown
in Table~\ref{rates}. Apart from the total event rates,
Super--Kamiokande has also measured the dependence of the event rates
during the day and during the night and the electron recoil energy
spectrum, all measured with their recent 1117-day data
sample~\cite{sksol00}. Although, as discuss in
Ref.~\cite{3solmsw,3solvac,nu2000,four} the inclusion of Kamiokande results
does not affect the shape of the regions, because of the much larger
precision of the Super--Kamiokande data, it is convenient to introduce
it as in this way the number of degrees of freedom (d.o.f.) for the
rates--only--fit is $4-3=1$ (instead of zero), thus
allowing for the determination of a well--defined $\chi^2_{min}$
confidence level (CL).
 
For the calculation of the theoretical expectations we use the BP98
standard solar model of Ref.~\cite{bp98}.  The general expression of
the expected event rate in the presence of oscillations in experiment
$i$ in the three--neutrino framework is given by $R^{th}_i$ :
\begin{eqnarray} 
R^{th}_i & = & \sum_{k=1,8} \phi_k 
\int\! dE_\nu\, \lambda_k (E_\nu) \times  
\Big[ \sigma_{e,i}(E_\nu)  \langle  
P_{\nu_e\to\nu_e}\rangle \label{ratesth} \\ 
& &                            + \sigma_{x,i}(E_\nu)  
\bigg(1-\langle P_{\nu_e\to\nu_e}\rangle \bigg)\Big] \nonumber. 
\end{eqnarray}   
where $E_\nu$ is the neutrino energy, $\phi_k$ is the total neutrino
flux and $\lambda_k$ is the neutrino energy spectrum (normalized to 1)
from the solar nuclear reaction $k$ with the normalization given in
Ref.~\cite{bp98}. Here $\sigma_{e,i}$ ($\sigma_{x,i}$) is the $\nu_e$
($\nu_x$, $x=\mu,\,\tau$) interaction cross section in the SM with the
target corresponding to experiment $i$.  For the Chlorine and Gallium
experiments we use improved cross sections $\sigma_{e,i}(E)$ from
Ref.~\cite{prod}. For the Kamiokande and Super--Kamiokande experiment
we calculate the expected signal with the corrected cross section as
explained below. Finally $\langle P_{\nu_e\to\nu_\alpha} \rangle$ is
the time--averaged $\nu_e$ survival probability in Eq.~(\ref{Peesol}).

We have also included in the fit the experimental results from the
Super--Kamiokande Collaboration on the day--night variation of the
event rates and the recoil electron energy spectrum.  In previous
works~\cite{2solour,2solother} the data on the zenith angular
dependence taken on 5 night periods and the day averaged value, and
the daily average recoil energy spectrum were included in order to
statistically combine the information on the day--night variation and
the energy dependence.  In principle, such analysis should be taken
with a grain of salt as these pieces of information are not fully
independent; in fact, they are just different projections of the
double differential spectrum of events as a function of time and
energy and can be subject to possible correlations between the
uncertainties in the energy and time dependence of the event rates,
which are neglected.  Here, instead, we follow the analysis of
Ref.~\cite{nu2000} and, in order to combine both the day--night
information and the spectral data we use the separately measured
recoil electron energy spectrum during the day and during the night
which is free of the unknown correlated uncertainties as they
correspond to different data samples. This will be referred in the
following as the day--night spectra data which contains $2\times 18$
data bins, including the results from the LE analysis for the 16 bins
above 6.5 MeV and the results from the SLE analysis for the two low
energy bins below 6.5 MeV.

The general expression of the expected rate in the bin $j$ during the Day 
(Night) in the presence of oscillations, $R_{sk,j}^{th,D(N)}$  
is similar to that in Eq.~(\ref{ratesth}), 
with the substitution of the cross sections with the corresponding 
differential cross sections 
folded with the finite energy resolution function of the detector 
and integrated over the electron recoil energy interval of the bin, 
$T_{\text {min}}\leq T\leq T_{\text {max}}$: 
\begin{equation} 
\sigma_{\alpha,sk}(E_\nu)=\int_{T_{\text {min}}}^{T_{\text {max}}}\!dT 
\int_0^{\frac{E_\nu}{1+m_e/2E_\nu}} 
\!dT'\,Res(T,\,T')\,\frac{d\sigma_{\alpha,sk}(E_\nu,\,T')}{dT'}\ . 
\label{sigma} 
\end{equation} 
The resolution function $Res(T,\,T')$ is of the form~\cite{sksol,Faid}: 
\begin{equation} 
Res(T,\,T') = \frac{1}{\sqrt{2\pi}(0.47 \sqrt{T'\text{(MeV)}})}\exp 
\left[-\frac{(T-T')^2}{0.44\,T' ({\text {MeV}})}\right]\ , 
\end{equation} 
and we take the differential cross section $d\sigma_\alpha(E_\nu,\,T')/dT'$  
from~\cite{CrSe}. When computing the spectrum during the day no Earth 
regeneration effect is included in the computation of 
$P_{\nu_e\to\nu_\alpha}$  while during the night such effect is included
as described in Sec.~\ref{probs}.

In the statistical treatment of all these data we perform a $\chi^2$ 
analysis for the different sets of data and  
we define a $\chi^2$ function for the set of observables  
$\chi^2_{\odot,\text {rates}}$ and 
$\chi^2_{\odot,\text {spec DN}}$. For the rates we   
follow closely the analysis of Ref.~\cite{fogli-lisi} with the  
updated uncertainties given in Refs.~\cite{bp98,prod}, as 
discussed in Ref.~\cite{2solour,nu2000}. For the day--night spectra we 
adopt a definition following Ref.~\cite{2solour}:
\begin{equation}
\chi^2_{\odot,\text{spec DN}}=\sum_{D,N} \sum_{i,j=1,18} 
(\alpha_{sp,dn}\frac{\displaystyle R^{th}_i}{R^{\rm BP98}_i} 
-R^{exp}_i)
\sigma_{ij}^{-2} (\alpha_{sp}\frac{\displaystyle R^{th}_j}{R^{\rm BP98}_j}
 - R^{exp}_j)
\end{equation}
where 
\begin{equation}
\sigma^2_{ij}=\delta_{ij}(\sigma^2_{i,stat}+\sigma^2_{i,uncorr})+
\sigma_{i,exp} \sigma_{j,exp}+\sigma_{i,cal}\sigma_{j,cal}
\end{equation}
describes the correlated and uncorrelated errors included in the
Super--Kamiokande spectra described in Refs.~\cite{2solour}. Notice
that in $\chi^2_{\odot,\text {spec DN}}$ we allow for a free
normalization in order to avoid double-counting with the data on the
total event rate which is already included in $\chi^2_{\odot,\text
  {rates}}$.  In the combinations of observables we define the
$\chi^2$ of the combination as the sum of the two $\chi^2$'s.  As
discussed in Sec.~\ref{probs} for the analysis of solar neutrino data
the oscillation probabilities depend only on three parameters:
$\theta_{12}$, $\theta_{13}$ and $\Delta m^2_{21}$.  Minimizing
$\chi^2_{\odot, \text{OBS}}$ for a given combination, OBS, of
solar neutrino experiments as a function of the three neutrino
oscillation parameters we determine their best fit value as well as
the corresponding $90\,(95)\,[99]\,$\% CL allowed regions for three
degrees of freedom by the condition
\begin{equation} 
\chi^2_{\odot, \text{OBS}} (\Delta m^2_{21},\theta_{12},\theta_{13})
-\chi^2_{min, \odot, \text{OBS}}\leq \Delta\chi^2 \mbox{(CL, 3~d.o.f.)} 
\label{delchi3} 
\end{equation}
where, for instance, $\Delta\chi^2\text{(CL, 3~d.o.f.)} = 6.25$, 7.81, and
11.34 for $\text{CL} = 90$, 95, and 99\% respectively.

\subsection{Atmospheric Neutrinos}
\label{data:atm}

Underground experiments can record atmospheric neutrinos by direct
observation of their charged current interaction inside the detector.
These so-called contained events, can be further classified into fully
contained events, when the charged lepton (either electron or muon)
produced by the neutrino interaction does not escape the detector, and
partially contained muons when the latter, produced inside, leaves the
detector. For Kamiokande and Super--Kamiokande, the contained data
sample is further divided into sub-GeV events with visible energy
below 1.2~GeV, and multi-GeV events, with lepton energy above this
cutoff. Sub-GeV events arise from neutrinos of several hundreds of
MeV, while multi-GeV events are originated by neutrinos with energies
of several GeV.

Contained events have been recorded at six underground experiments,
using water-Cerenkov detectors: Kamiokande~\cite{Kamiokande},
IMB~\cite{IMB} and Super--Kamiokande~\cite{skcont,skatm00}, as well as
iron calorimeters: Fr\'ejus~\cite{Frejus}, NUSEX~\cite{Nusex} and
Soudan2~\cite{Soudan2}.
The expected number of $e$-like and $\mu$-like contained events,
$N_\beta$ ($\beta = e,\mu$), is given as
\begin{equation} \label{eq:event0}
    N_\beta = n_t T \int \sum_{\alpha,\pm}
    \frac{d^2 \Phi_\alpha^\pm}{dE_\nu \, dc_\nu} \kappa_\alpha(h,c_\nu,E_\nu)
    P_{\alpha\beta}^\pm \frac{d\sigma_\beta^\pm}{dE_l}
    \varepsilon_\beta(E_l) \, dE_\nu \, dE_l \, dc_\nu \, dh
\end{equation}
where $P_{\alpha\beta}^+$ ($P_{\alpha\beta}^-$) is the $\nu_\alpha \to
\nu_\beta$ ($\bar{\nu}_\alpha \to \bar{\nu}_\beta$) conversion
probability for given values of the neutrino energy $E_\nu$, the
cosine $c_\nu$ of the angle between the incoming neutrino and the
vertical direction, and the slant distance $h$ from the production
point to the sea level. In the SM one has $P_{\alpha\beta}^\pm =
\delta_{\alpha\beta}$ for all $\alpha,\beta$.  In
Eq.~(\ref{eq:event0}) $n_t$ is the number of targets, $T$ is the
experiment running time and $\Phi_\alpha^+$ ($\Phi_\alpha^-$) is the
flux of atmospheric neutrinos (antineutrinos) of type $\alpha=e,\mu$,
for which we will adopt the Bartol flux~\cite{fluxes1}; $E_l$ is the
energy of the final charged lepton of type $\beta=e,\mu$,
$\varepsilon_\beta(E_l)$ is the detection efficiency for such lepton
and $\sigma_\beta^+$ ($\sigma_\beta^-$) is the neutrino-
(antineutrino-) nucleon interaction cross section.
Finally, $\kappa_\alpha$ is the slant distance distribution,
normalized to one~\cite{pathlength}. For the angular distribution of
events we integrate in the corresponding bins in $c_l \equiv
\cos\theta_l$ where $\theta_l$ is the angle of the detected lepton,
taking into account the opening angle between the neutrino and the
charged lepton directions as determined by the kinematics of the
neutrino interaction. On average the angle between the final-state
lepton and the incoming neutrino directions ranges from $70^\circ$ at
200~MeV to $20^\circ$ at 1.5~GeV.
One must also take into consideration that the neutrino fluxes,
especially in the sub-GeV range, depend on the solar activity. In
order to take this fact into account, we use in Eq.~(\ref{eq:event0})
a linear combination of atmospheric neutrino fluxes
$\Phi_\alpha^{\pm,max}$ and $\Phi_\alpha^{\pm,min}$ which correspond
to the most active Sun (solar maximum) and quiet Sun (solar minimum),
respectively, with different weights depending on the running period
of each experiment~\cite{2atmour}.  The agreement of our predictions
with the experimental Monte Carlo predictions was explicitly verified
in Ref.~\cite{2atmour}.  This renders confidence in the reliability of
our results for contained events.

Higher energy muon neutrinos and anti-neutrinos are detected
indirectly by observing the muons produced by charged current
interactions in the vicinity of the detector: the so called upgoing
muons. If the muon stops inside the detector, it will be called a
``stopping'' muon, while if the muon track crosses the full detector
the event is classified as a ``through--going'' muon.  On average
stopping muons arise from neutrinos with energies around ten GeV,
while through--going muons are originated by neutrinos with energies
around hundred GeV.  In our analysis we will consider the latest
results from Super--Kamiokande~\cite{skatm00} and from the
MACRO~\cite{MACRO} experiment\footnote{We decided not to
  include Baksan~\cite{Baksan} data because they appear
  inconclusive.} on upgoing muons which are presented in the form of
measured muon fluxes. We obtain the effective muon fluxes for both
stopping and through--going muons by convoluting the $\nu_\alpha \to
\nu_\mu$ transition probability (calculated as in Sec.~\ref{probs})
with the corresponding muon fluxes produced by the neutrino
interactions with the Earth. We include the muon energy loss during
propagation both in the rock and in the detector according to
Refs.~\cite{muloss,ricardo} taking into account also the effective
detector area for stopping and through--going events.  Schematically
\begin{equation}
    \Phi_\mu(c_\nu)_{S,T}=\frac{1}{A(L_{min},c_\nu)}
    \int_{E_{\mu,min}}^{\infty}
    \frac{d^2\Phi_\mu(E_\mu,c_\nu)}{dE_\mu dc_\nu}
    A_{S,T}(E_\mu,c_\nu)dE_{\mu} \, \; ,
    \label{eq:upmuons1}
\end{equation}  
where
\begin{eqnarray}
    \frac{d^2\Phi_\mu}{dE_\mu dc_\nu}
    &=& N_A \int_{E_{\mu}}^\infty dE_{\mu 0}
    \int_{E_{\mu 0}}^\infty dE_\nu
    \int_0^\infty dX  \int_0^\infty dh  \nonumber \\
    & & \sum_{\alpha,\pm}
    \frac{d^2 \Phi_\alpha^\pm}{dE_\nu \, dc_\nu} 
    \kappa_\alpha(h,c_\nu,E_\nu) P_{\alpha\mu}^\pm 
    \frac{d\sigma_\mu^\pm(E_\nu,E_{\mu 0})}{dE_{\mu 0}} \,
    F_{rock}(E_{\mu 0}, E_\mu, X)
    \label{eq:upmuons2}
\end{eqnarray}
where $N_A$ is the Avogadro number, $E_{\mu 0}$ is the energy of the
muon produced in the neutrino interaction and $E_\mu$ is the muon
energy when entering the detector after travelling a distance $X$ in
the rock. At the relevant energies the opening angle between incident
neutrino and outgoing muon can be neglected to a very good
approximation, thus we use a common label $c_\nu$ to characterize both
directions. Here $F_{rock}(E_{\mu 0}, E_\mu, X)$ is the function which
characterizes the energy spectrum of the muons arriving the detector.

In Eq.~(\ref{eq:upmuons1}) $A(L_{min},c_\nu) = A_{S}(E_\mu,c_\nu) +
A_{T}(E_\mu,c_\nu)$ is the projected detector area for internal
path-lengths longer than a certain $L_{min}$. Here $A_{S}$ and $A_{T}$
are the corresponding effective areas for stopping and through--going
muon trajectories. For Super--Kamiokande $L_{min} = 7$~m and
we compute these effective areas using the simple
geometrical picture given in Ref.~\cite{lipari1}.

In contrast with Super--Kamiokande, MACRO presents its results as muon
fluxes for $E_\mu>1$~GeV, after correcting for detector acceptances.
Therefore in this case we compute the expected fluxes as in
Eqs.~(\ref{eq:upmuons1}) and~(\ref{eq:upmuons2}) but without the
inclusion of the effective areas.  In Ref.~\cite{2atmour} we have
explicitly verified that our predictions for upgoing muons agree with
the experimental Monte Carlo predictions from Super--Kamiokande and
MACRO to the 5\% and 1\% level, respectively.

For the statistical treatment of all these data we perform a $\chi^2$
analysis for different sets of data by computing the
$\chi^2_{atm,\text{OBS}}$ for a given combination, OBS, of
experiments as a function of the neutrino oscillation parameters.
Following closely the analysis of Refs.~\cite{2atmour,3atm} we use the
previously described contained and upgoing event numbers (instead of
their ratios), paying attention to the correlations between the
sources of errors in the muon and electron predictions, as well as the
correlations amongst the errors of different energy data samples.
Thus we define $\chi^2_{atm,\text{OBS}}$ as
\begin{equation} \label{eq:chi2}
    \chi^2_{atm,\text{OBS}} \equiv \sum_{I,J\in {\text{OBS}}}
    (N_I^{DA}-N_I^{TH}) \cdot 
    (\sigma_{DA}^2 + \sigma_{TH}^2 )_{IJ}^{-1} \cdot
    (N_J^{DA}-N_J^{TH})\;,
\end{equation}
with 
\begin{equation}
    \left[\sigma_{DA(TH)}^2\right]_{IJ} \equiv 
   \sigma_{DA(TH),\alpha}(A) \, \rho_{DA(TH),\alpha \beta} (A,B) \,
    \sigma_{DA(TH),\beta}(B).
\end{equation}
$I$ and $J$ stand for any combination of experimental data sets and
event-types considered, {\it i.e}, $I = (A, \alpha)$ and $J =
(B,\beta)$. The latin indices $A, B$ stand for the different
experiments or different data samples in a given experiment. The greek
indexes denote $e$-- or $\mu$--type events, {\it i.e}, $\alpha,
\beta = e, \mu$. Here $N_I^{TH}$ denotes the predicted number of
events (or the predicted value of the flux, in the case of upgoing
muons) calculated as discussed above, whereas $N_I^{DA}$ is the
corresponding experimental measurement.  We denote by
$\sigma_{DA(TH)}^2$ the error matrices containing the experimental
(theoretical) errors and by $\rho_{DA(TH) \alpha \beta} (A,B)$ the
matrix containing all the correlations between the experimental
(theoretical) errors of $\alpha$-like events in the $A$ experiment and
$\beta$-like events in $B$ experiment, whereas
$\sigma_{DA(TH),\alpha}(A)$ is the experimental (theoretical) error
for the number of events in the $A$ experiment.  The dimensionality of
the error matrix depends on the combination of experiments, OBS,
included in the analysis.  A detailed discussion of the errors and
correlations used in our analysis can be found in the appendixes of
Refs.~\cite{2atmour} both for contained and for the upgoing muon data
analysis.

As discussed in Sec.~\ref{probs} for the analysis of any set of
atmospheric neutrino data the oscillation probabilities depend only on
three parameters: $\theta_{13}$, $\theta_{23}$ and $\Delta m^2_{32}$.
Minimizing $\chi^2_{atm,\text{OBS}}$ with respect to these three
parameters we determine their best fit value as well as the
corresponding $90\,(95)\,[99]\,$\% confidence levels (CL) allowed
regions for three degrees of freedom by the conditions in
Eq.~(\ref{delchi3}) where now $\chi^2_{atm,\text{OBS}}(\Delta
m^2_{32},\theta_{23},\theta_{12})$.

\subsection{Reactor Neutrinos: CHOOZ}
\label{data:CHOOZ}

The CHOOZ experiment~\cite{CHOOZ} searches for disappearance of
$\bar{\nu}_e$ produced in a power station with two pressurized-water
nuclear reactors with a total thermal power of $8.5 \,GW$ (thermal).
At the detector, located at $L\simeq 1$ Km from the reactors, the
$\bar{\nu}_e$ reaction signature is the delayed coincidence between
the prompt ${\rm e^+}$ signal and the signal due to the neutron
capture in the Gd-loaded scintillator.  Their measured vs. expected
ratio, averaged over the neutrino energy spectrum is
\begin{equation}
R = 1.01 \pm 2.8 \,\% ({\rm stat}) \pm 2.7 \,\% ({\rm syst})
\label{rchooz}
\end{equation}
Thus no evidence was found for a deficit of measured vs. expected
neutrino interactions and they derive from the data exclusion plots in
the plane of the oscillation parameters $(\Delta m^2,\sin^2 2
\theta)$, in the simple two-neutrino oscillation scheme. At 90\% CL
they exclude the region given by approximately $\Delta m^2 > 7 \cdot
10^{-4}$ eV$^2$ for maximum mixing, and $\sin^2(2\theta) = 0.10$ for
large $\Delta m^2$.

\begin{figure} 
\begin{center} 
\mbox{\epsfig{file=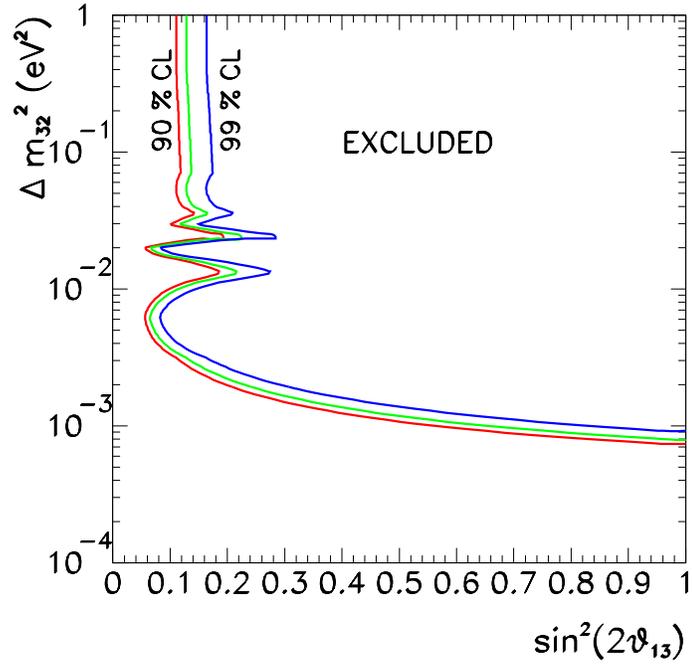,height=0.4\textheight}} 
\end{center} 
\caption{Excluded region in  $\Delta{m}^2_{32}$ and $\sin^2(2\theta_{13})$  
from the non observation of oscillations by the CHOOZ reactor experiment.
The curves represent the 90, 95 and 99\% CL excluded region 
defined with 2 d.o.f.\ for comparison with the CHOOZ
published results.} 
\label{fig:chooz} 
\end{figure}

In order to combine the CHOOZ results with the results from our
analysis of solar and atmospheric neutrino data in the framework of
three--neutrino mixing we have first performed our own analysis of the
CHOOZ data. Using as experimental input their measured ratio
(\ref{rchooz})~\cite{CHOOZ} and comparing it with the theoretical
expectations we define the $\chi^2_{\text{CHOOZ}}$ function.  
As discussed in Sec.~\ref{probs} for the analysis of the reactor data the
relevant oscillation probability depends in general on four parameters
$\theta_{12}$, $\Delta m^2_{21}$, $\theta_{13}$, and $\Delta
m^2_{32}$, but for $\Delta m^2_{21}\lesssim 3\times 10^{-4}$ eV$^2$, 
even in the three--neutrino mixing scenario, it only depends on the last 
two. We verified that in this case with our 
$\chi^2_{\text{CHOOZ}}$ function
and using the statistical criteria for two degrees of freedom we
reproduce the excluded regions given in Ref.~\cite{CHOOZ} for
two--neutrino oscillations as can be seen in Fig.~\ref{fig:chooz}
where we show the excluded regions at 90, 95 and 99\% CL in the
$\left(\Delta m^2_{32},\sin^2(2\theta_{13})\right)$ plane from our
analysis of the CHOOZ data defined with 2~d.o.f.\
($\Delta\chi^2_{\text{CHOOZ}}=4.61$, 6.0, 9.21 respectively). Comparing
our results with Ref.~\cite{CHOOZ} we find a very good agreement.
 
\section{Three--neutrino Oscillation Analysis of Solar Data}
\label{analysis:solar}

As explained in Sec.~\ref{probs}, for the mass scales involved in the
explanation of the solar and atmospheric data the relevant parameter
space for solar neutrino oscillations in the framework of
three--neutrino mixing is a three dimensional space in the variables
$\Delta{m}^2_{21}$, $\theta_{12}$ and $\theta_{13}$.  In our choice of
ordering for the neutrino masses the mass-squared difference
$\Delta{m}^2_{21}$ is positive and the mixing angles $\theta_{12}$ and
$\theta_{13}$ can vary in the interval $0 \leq \theta_{12} \leq
\frac{\pi}{2}$ and $0 \leq \theta_{13} \leq \frac{\pi}{2}$.  In our
analysis we choose to parametrize the $\theta_{12}$ and $\theta_{13}$
dependence in terms of the $\tan^2\theta_{12}$ and $\tan^2\theta_{13}$
variables which span the full parameter space.

\begin{figure} 
\begin{center} 
\mbox{\epsfig{file=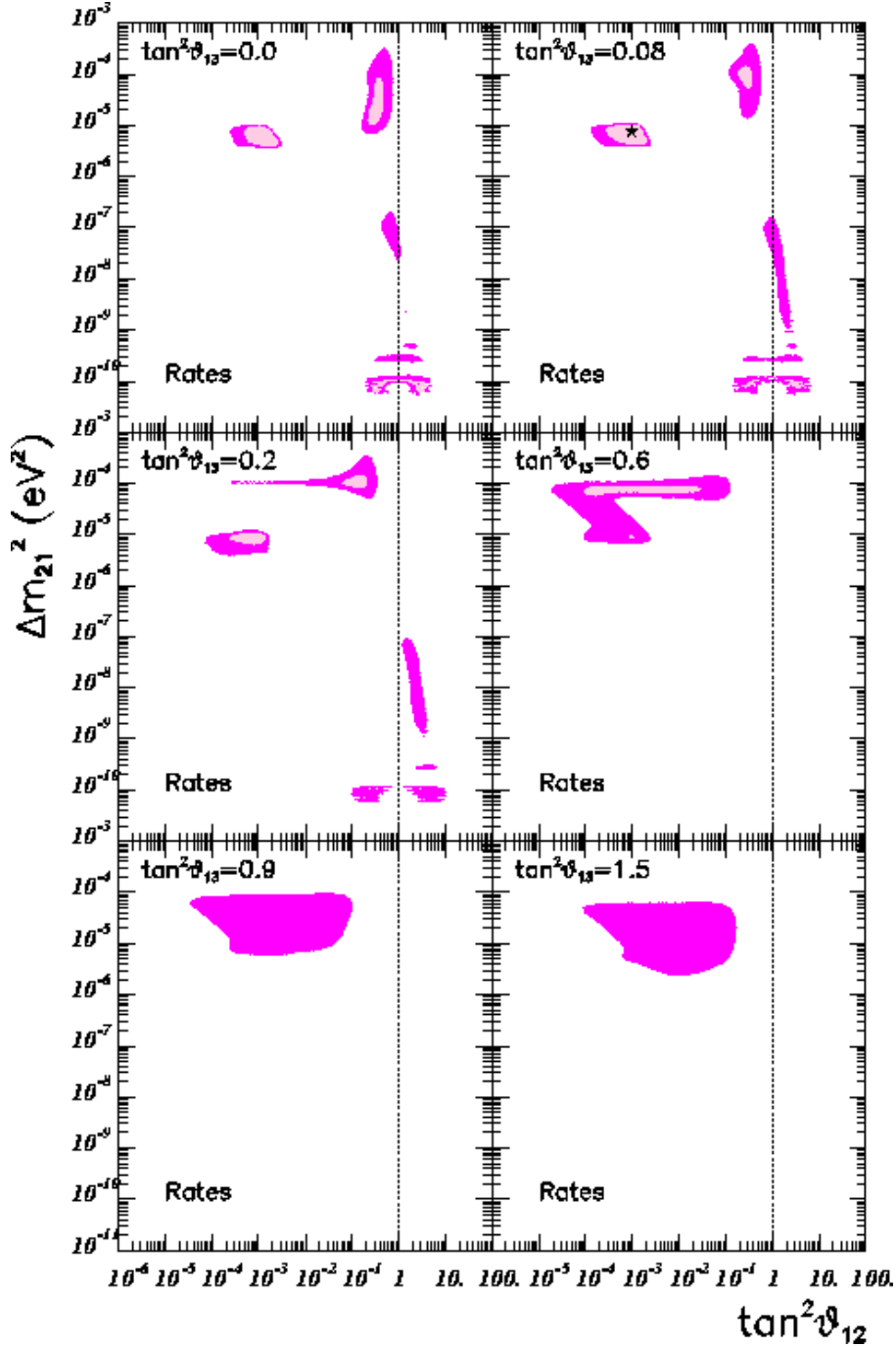,height=0.8\textheight}} 
\end{center} 
\caption{Allowed three--neutrino oscillation regions in 
  $\Delta{m}^2_{21}$ and $\tan^2\theta_{12}$ from the measurements of
  the total event rates at Chlorine, Gallium, Kamiokande and
  Super--Kamiokande (1117-day data sample) experiments. The different
  panels represent the allowed regions at 99\% (darker) and 90\% CL
  (lighter) obtained as sections for fixed values of the mixing angle
  $\tan^2\theta_{13}$ of the three--dimensional volume defined by
  $\chi^2-\chi^2_{min}$=6.25 (90\%), 11.36 (99\%). The best--fit point is 
denoted as a star.}
\label{fig:3sol_r} 
\end{figure}

\begin{figure} 
\begin{center} 
\mbox{\epsfig{file=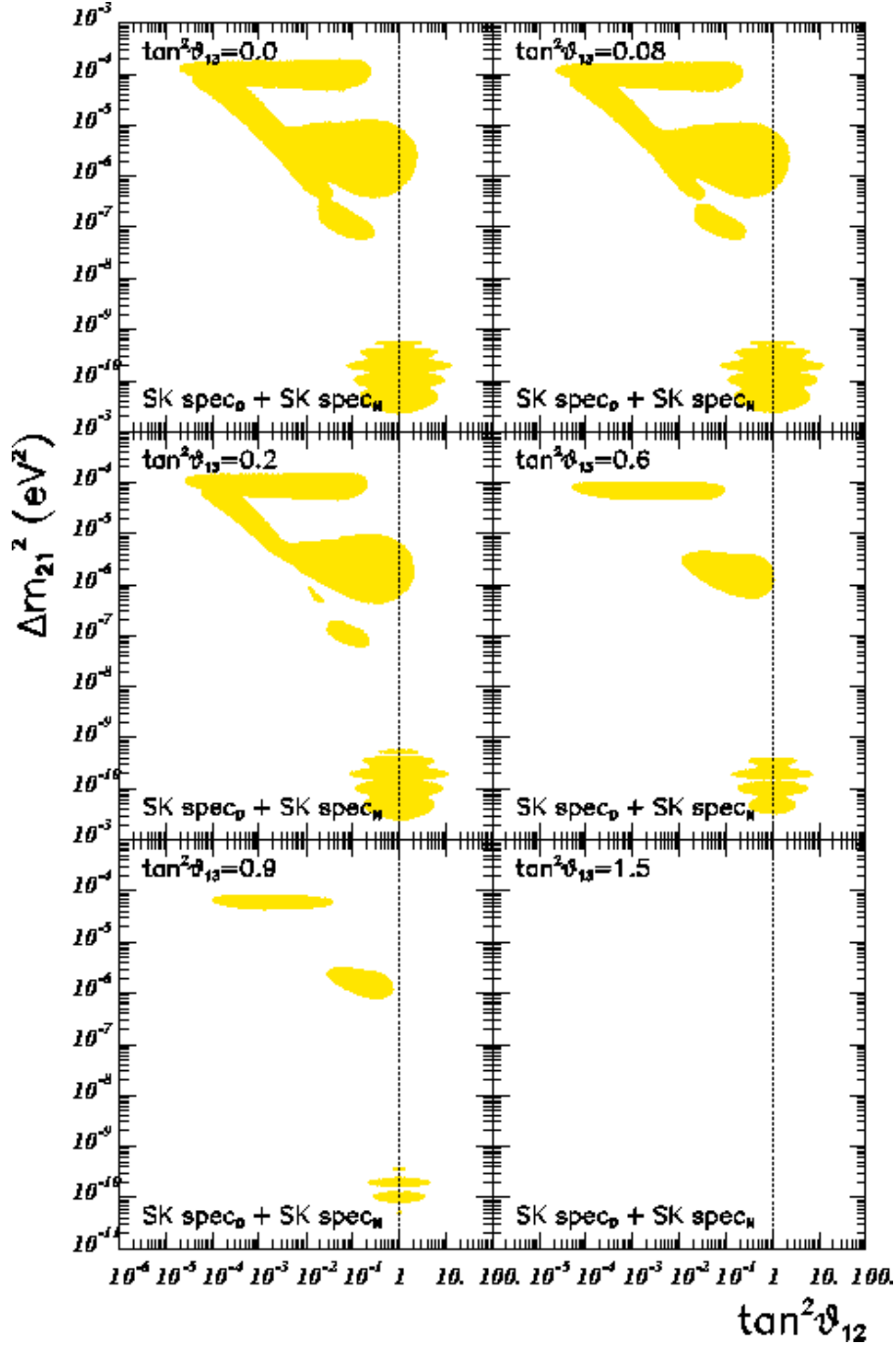,height=0.8\textheight}} 
\end{center} 
\caption{Excluded three--neutrino solar oscillation regions 
  at 99\% CL in $\Delta{m}^2_{21}$ and $\tan^2\theta_{12}$ from the
  measurement of the day--night spectra data by Super--Kamiokande
  (1117-day data sample).  }
\label{fig:3sol_ex} 
\end{figure}

\begin{figure} 
\begin{center} 
\mbox{\epsfig{file=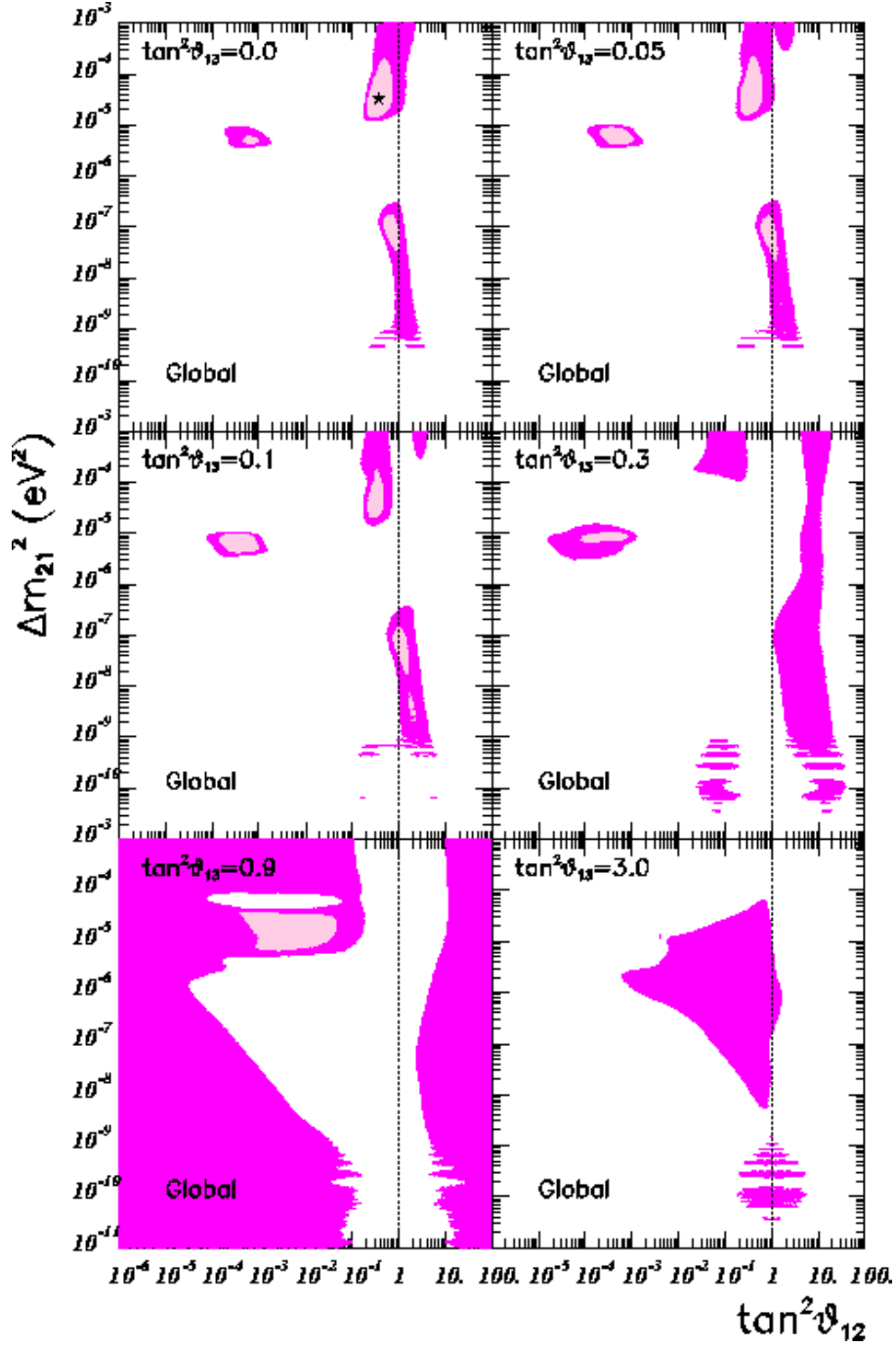,height=0.8\textheight}} 
\end{center} 
\caption{Allowed three--neutrino solar oscillation regions in  
  $\Delta{m}^2_{21}$ and $\tan^2\theta_{12}$ from the global analysis
  of solar neutrino data. The best--fit point is denoted as a star.}
\label{fig:3sol_rspdn} 
\end{figure}

We first present the results of the allowed regions in the
three--parameter space for the different combination of observables.
In building these regions, for a given set of observables, we compute
for any point in the parameter space of three--neutrino oscillations
the expected values of the observables and with those and the
corresponding uncertainties we construct the function
$\chi^2_{\odot,\text{obs}}(\Delta m^2_{21},\theta_{12},\theta_{13})$.
We find its minimum in the full three-dimensional space considering
both MSW and vacuum oscillations as well as the transition regime of
quasi--vacuum oscillations on the same footing. The allowed regions
for a given CL are then defined as the set of points satisfying the
conditions given in Eq.~(\ref{delchi3}). In
Figs.~\ref{fig:3sol_r}--\ref{fig:3sol_rspdn} we plot the sections of
such volume in the plane ($\Delta{m}^2_{21},\tan^2(\theta_{12})$) for
different values of $\tan^2\theta_{13}$.
 
\begin{figure} 
\begin{center} 
\mbox{\epsfig{file=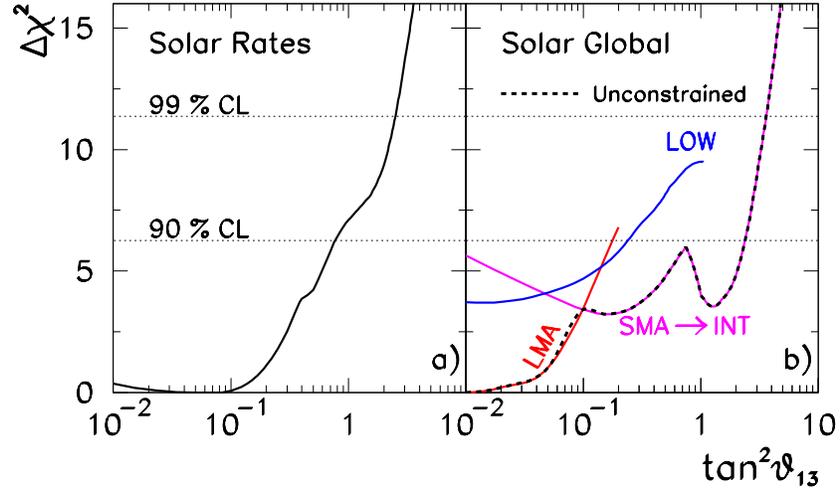,width=0.7\textwidth}} 
\end{center} 
\caption{$\Delta \chi^2$ as a function of $\tan^2\theta_{13}$
  from the three--neutrino analysis of the solar data.  The left panel
  corresponds the analysis of total rates only and the right panel to
  the global analysis.  The dotted horizontal lines correspond to the
  90\%, 99\% CL limits.}
\label{fig:chisol} 
\end{figure}

Figure~\ref{fig:3sol_r} shows the results of the fit to the observed
total rates only. For the sake of clarity we show the regions only at
90 and 99\% CL.  We find that for small $\tan^2\theta_{13} \lesssim
0.3$ both at 90 and 99\% CL, the three--dimensional allowed volume is
composed of four separated three-dimensional regions in the MSW sector
of the parameter space which we denote as SMA, LMA and LOW solutions
analogous to the usual two--neutrino oscillation picture and a
``tower'' of regions in the vacuum oscillations (VO) sector.  The
global minimum $\chi^2_{min,\odot,\text{rates}}=0.62$ (see
Table~\ref{table:chisol}) used in the construction of the volumes lies
in the SMA region and for a non--vanishing value of $\tan^2\theta_{13}
\simeq 0.07$. However, as can be seen in Fig.~\ref{fig:chisol}, this
has hardly any statistical significance, as $\Delta\chi^2$ is very
mildly dependent on $\theta_{13}$ for these small values of
$\tan^2\theta_{13}$.

As seen in Fig.~\ref{fig:3sol_r} as $\tan^2\theta_{13}$ grows we find
the following behaviours of the allowed regions
\begin{itemize}
\item [$(i)$] For small values of $\theta_{13}$ the SMA region
  migrates towards larger $\Delta{m}^2_{21}$ and the LMA region
  migrates towards lower mixing angles and larger $\Delta{m}^2_{21}$.
  The increase of $\theta_{13}$ produces that the SMA and LMA regions
  merge into a unique allowed intermediate (INT) region which
  disappears at larger values of $\tan^2\theta_{13}$
\item [$(ii)$] The LOW region migrates towards the second octant and
  lower $\Delta{m}^2_{21}$ and then disappears
\item [$(iii)$] The 99\% CL region for vacuum solution first grow for
  small values of $\theta_{13}$ and then become smaller as
  $\tan^2\theta_{13}$ increases until it finally disappears.
\end{itemize}
$(i)$ and $(ii)$ are in agreement with the results of
Ref.~\cite{3solmsw} while $(iii)$ agrees with the results of the
second reference in~\cite{3solvac}.

Thus from Fig.~\ref{fig:3sol_r} we find that as $\tan^2\theta_{13}$
increases all the allowed regions from the fit to the total event
rates disappear, leading to an upper bound on $\tan^2\theta_{13}$ for
any value of $\Delta m^2_{21}$, independently of the values taken by
the other parameters in the three--neutrino mixing matrix.  In
Fig.~\ref{fig:chisol}.a we plot
$\Delta\chi^2_{\odot,\text{rates}}(\tan^2\theta_{13})=
\chi^2_{\odot,\text{rates}}(\tan^2\theta_{13})
-\chi^2_{min,\odot,\text{rates}}$ where
$\chi^2_{\odot,\text{rates}}(\tan^2\theta_{13})$ is obtained by
minimizing $\chi^2_{\odot,\text{rates}}(\Delta m^2_{21},
\theta_{12},\theta_{13})$ with respect to $\Delta m^2_{21}$ and
$\theta_{12}$ for fixed values of $\theta_{13}$ and where
$\chi^2_{min,\odot,\text{rates}}=0.62$ is the global minimum in the
full three parameter space. From the figure we can extract the upper
limit on $\tan^2\theta_{13}$ from the analysis of the total event
rates. The corresponding 90 and 99\% CL bounds are tabulated in
Table~\ref{table:limits}.

Figure~\ref{fig:3sol_ex} shows the region excluded at 99\% CL by the
Super--Kamiokande day--night spectra data for the same
$\tan^2\theta_{13}$ values as in Fig.~\ref{fig:3sol_r}.  The global
minimum $\chi^2_{min,\odot,\text {spec DN}}=28.0$ (see
Table~\ref{table:chisol}) used in the construction of the volumes lies
in the VO region and corresponds to $\tan^2\theta_{13}=0$.  As seen
in the figure for $\tan^2\theta_{13}=0$ the excluded region overlaps
with the lower part of the LMA allowed region (where a larger Earth
regeneration effect is expected), covers a big fraction of the SMA
region and practically the full VO region (where larger distortion of
the energy spectrum is expected).  As $\tan^2\theta_{13}$ increases the
excluded region becomes smaller.  This arises from the fact that in
the survival probability (see Eq.~(\ref{Peesol})) the energy
independent term $\sin^4\theta_{13}$ increases and the flat recoil
electron energy spectrum can be more easily accommodated. For
intermediate $\tan^2\theta_{13}$ values (see third and fourth panels
in Figs.~\ref{fig:3sol_r} and~\ref{fig:3sol_ex} ) there is a large
overlap of the excluded region with the INT region where the LMA and
SMA merge.

In Fig.~\ref{fig:3sol_rspdn} we show the results from the global fit
of the full solar data--set including the total observed rates and the
Super--Kamiokande day--night spectra data.  The global minimum
$\chi^2_{min,\odot,\text{global}}=35.2$ (see Table~\ref{table:chisol})
used in the construction of the volumes lies in the LMA region and
corresponds to $\tan^2\theta_{13}=0$.  The behaviour of the regions
illustrate the ``tension'' between the data on the total event rates
which favour smaller $\theta_{13}$ values and the day--night spectra
which allow larger values. It can also be understood
as the ``tension'' between the energy dependent and constant pieces of
the electron survival probability in Eq.~(\ref{Peesol}).
As a consequence of this ``tension'' between the two behaviours the
bound on $\theta_{13}$ from the global analysis, which we list in
Table~\ref{table:limits}, is weaker than the one derived from the
analysis of the event rates only. One may wonder about the meaning of
the ``allowed regions'' for such large values of $\theta_{13}$. To
clarify this point we have defined the following ``sectors'' in the
($\Delta m^2_{21},\tan^2\theta_{12}$) plane
\begin{itemize}
\item{LMA:}
\begin{eqnarray}
 1\times 10^{-3}\; & <\; \Delta m^2_{21}/\mbox{\rm eV$^2$}\;< \;& 
\;1\times 10^{-5}\;\,\\\nonumber
 1\times 10^{-1} & <\;\tan^2\theta_{12}\;< \;& 
\;10\,
\label{solarrangeslma}
\end{eqnarray}
\item{SMA:}
\begin{eqnarray}
 1\times 10^{-4}\; & <\; \Delta m^2_{21}/\mbox{\rm eV$^2$}\;< \;& 
\;1\times 10^{-6}\;\,\\\nonumber
 1\times 10^{-5} & <\;\tan^2\theta_{12}\;< \;& \;1\times 10^{-3}\,
\label{solarrangessma}
\end{eqnarray}
\item{LOW:}
\begin{eqnarray}
 1\times 10^{-8}\; & <\; \Delta m^2_{21}/\mbox{\rm eV$^2$}\;< \;& 
\;1\times 10^{-6}\;\,\\\nonumber
 1\times 10^{-1} & <\;\tan^2\theta_{12}\,
\label{solarrangeslow}
\end{eqnarray}
\end{itemize}
and we have studied the behaviour of the
$\chi^2_{\odot,\text{global}}$ in each of these sectors.  We find that
in each of these sectors there is a local minimum around which there
exists an allowed region if $\theta_{13}$ is below certain limit. As a
self-consistency check we have verified that those local minima are
well defined for any value of $\theta_{13}$ below the limit arising
from the analysis of atmospheric and/or reactor data (see
Sec.~\ref{analysis:atmos}).
 
In Fig.~\ref{fig:chisol}.b we plot the values of the
$\chi^2_{\odot,\text{global}}$ at the local minima as a function of
$\tan^2\theta_{13}$. The different curves correspond to the functions
\begin{displaymath}
\Delta\chi^2_{\odot,\text{global}}
(\tan^2\theta_{13}) |_{\text{sector}}=
\chi^2_{\odot,\text{global}}
(\tan^2\theta_{13})|_{\text{sector}}
-\chi^2_{min,\odot,\text{global}}
\end{displaymath} 
where $\chi^2_{\odot,\text{global}}=35.2$ is the global minimum in the
{\sl full} three--neutrino parameter space (see
Table~\ref{table:chisol}) and
$\chi^2_{\odot,\text{global}}(\tan^2\theta_{13})|_{\text{sector}}$ is
obtained my minimizing $\chi^2_{\odot,\text{global}}(\Delta m^2_{21},
\theta_{12},\theta_{13})$ with respect to $\Delta m^2_{21}$ and
$\theta_{12}$ in each of the above defined sectors for fixed values of
$\theta_{13}$.  In what follows we label as ``constrained'' the
results of analyses when the parameters $\Delta m^2_{21}$ and
$\theta_{12}$ are varied in a given sector while ``unconstrained''
refers to the case where we allow the variation of $\Delta m^2_{21}$
and $\theta_{12}$ in the full plane.  In Fig.~\ref{fig:chisol}.b the
curves for the constrained analyses are displayed up to the value of
$\theta_{13}$ which allwos the existence of the minimum in the
defined sectors.

In Fig.~\ref{fig:chisol}.b, we also plot the function
$\Delta\chi^2_{\odot,\text{global}}(\tan^2\theta_{13})=
\chi^2_{\odot,\text{global}}(\tan^2\theta_{13})
-\chi^2_{min,\odot,\text{global}}$ with
$\chi^2_{\odot,\text{global}}(\tan^2\theta_{13})$ obtained by
minimizing in the full $\Delta m^2_{21}$ and $\theta_{12}$ plane,
i.e., for the unconstrained fit. This curve is simply the lower
envolvent of the curves from the constrained analysis.  From the
figure we see that, unlike for the analysis of the rates only, the
function $\Delta\chi^2_{\odot,\text{global}}(\tan^2\theta_{13})$ when
the minimization is unconstrained, is not a monotonically growing
function but presents some local maxima and minima.  This behaviour is
due to the fact that the unconstrained minimum in the ($\Delta
m^2_{21}$, $\theta_{12}$) plane moves from one sector to another:
\begin{itemize}
\item For $\tan^2\theta_{13}<0.1$ 
the minimum lies in the LMA region. The 
best global fit point for LMA  corresponds to the  case
$\tan^2\theta_{13}=0$ where solar and atmospheric analyses decouple.  
\item  For $0.1<\tan^2\theta_{13}<0.75$ the minimum lies in the SMA region 
for which the local best fit point happens at 
$\tan^2\theta_{13}=0.16$. 
\item  At $\tan^2\theta_{13}>0.75$ the minimum moves from the SMA to 
the INT region (which contains the parameter space 
between the three regions defined above). 
The preferred $\tan^2\theta_{13}$  for this INT region is 
$\tan^2\theta_{13}=1.3$. 
\end{itemize}
For larger values of $\tan^2\theta_{13}$, a quick worsening of the
$\chi^2$ is produced due to the increase of the constant term in the
survival probability.

We can now describe more precisely the behaviour of the allowed
regions shown in Fig.~\ref{fig:3sol_rspdn}: for small
$\tan^2\theta_{13}$ the global fit excludes all the regions of the
oscillation ($\Delta m^2_{21},\tan^2\theta_{12}$) parameter plane where the
energy dependent piece of the survival probability in
Eq.~(\ref{Peesol}), $\cos^4\theta_{13} P_{e'e'}^{2\nu}$, is either too
little to account for the observed event rates or too large to account
for the flat spectrum. As $\theta_{13}$ increases the constant
$\sin^4\theta_{13}$ piece in the survival probability increases and as
a result the fit to the flat spectrum becomes good in the full
($\Delta m^2_{21},\tan^2\theta_{12}$) plane while the fit to the event rates
becomes better in those regions where the energy dependence of the
$P_{e'e'}^{2\nu}$ piece is stronger.  In this way for
$\tan^2\theta_{13}\gtrsim 0.4 $ the local minimum in the ($\Delta
m^2_{21},\tan^2\theta_{12}$) plane for fixed $\tan^2\theta_{13}$ moves from
the LMA region first to the SMA and finally to the INT region where
for low energy neutrinos the energy dependence of $P_{e'e'}^{2\nu}$ is
stronger. As $\tan^2\theta_{13}$ increases to much larger values the
survival probability becomes basically energy independent and the fit
to the event rates becomes too bad in the full parameter space.

Let us finally comment on the statistical meaning of the allowed
regions.  Notice that following the standard procedure the allowed
regions shown in Figs.~\ref{fig:3sol_r}--\ref{fig:3sol_rspdn} have
been defined in terms of shifts of the $\chi^2$ function for those
observables {\sl with respect to the global minimum}. Defined this
way, the size of a region depends on the {\sl relative} quality of its
local minimum with respect to the global minimum but from the size of
the region we cannot infer the actual {\sl absolute} quality of the
description in each region. That is given by the value of the
$\chi^2_{\odot,\text{global}}$ function at the local minimum (which
for this case we show in Fig.~\ref{fig:chisol}.b). From this analysis
we see that for small $\tan^2\theta_{13}$ the values of
$\chi^2_{\odot,\text{global}}$ at the local minimum in the different
regions are not so different.  For instance, for
$\tan^2\theta_{13}<0.2$ we find that the goodness of the fit (GOF) for
the different solutions is 55\% for the LMA and 37\% for the SMA and
LOW solutions\footnote{The small differences in the GOF with the
  results in Ref.~\cite{nu2000} are due to the effect of the
  additional $\theta_{13}$ parameter}.  Thus our conclusion is that
from the statistical point of view for small $\tan^2\theta_{13}$ all
solutions are acceptable since they all provide a reasonable GOF to
the full data set. Although LMA solution seem slightly favoured over
SMA and LOW solution these last two solutions cannot be ruled out at
any reasonable CL.
 
\section{Three--neutrino Oscillation Analysis of Atmospheric and Reactor Data}
\label{analysis:atmos}

\subsection{Atmospheric Neutrino Fit}

In our statistical analysis of the atmospheric neutrino events we use
the following data: (i) unbinned contained $e$-like and $\mu$-like
event rates from Frejus~\cite{Frejus}, IMB~\cite{IMB},
Nusex~\cite{Nusex}, Kamiokande sub-GeV~\cite{Kamiokande} and
Soudan2~\cite{Soudan2};  (ii) $e$-like and $\mu$-like data samples of
Kamiokande multi-GeV~\cite{Kamiokande} and Super--Kamiokande sub- and
multi-GeV~\cite{skatm00}, each given as a 5-bin zenith-angle
distribution\footnote{Note that for convenience and maximal
  statistical significance we prefer to bin the Super--Kamiokande
  contained event data in 5, instead of 10 bins.}; (iii)
Super--Kamiokande upgoing muon data including the stopping (5 bins in
zenith-angle) and through--going (10 angular bins) muon fluxes; (iv)
MACRO~\cite{MACRO} upgoing muons samples, with 10 angular bins.

In order to study the results for the different types of atmospheric
neutrino data we have defined the following combinations of data sets:
\begin{itemize}
\item{FINKS} The $e$-like and $\mu$-like event rates from the five 
experiments Fr\'ejus, IMB, Nusex, Kamiokande sub-GeV and Soudan-2.
It contains 10 data points.
\item{CONT--UNBIN} The rates in FINKS together with Kamiokande 
multi-GeV and Super--Kamiokande sub and multi-GeV $e$-like and $\mu$-like
event rates without including the angular information, which accounts
for a total of 16 data points.
\item{CONT--BIN} The rates in FINKS together with Kamiokande 
multi-GeV and Super--Kamiokande sub and multi-GeV $e$-like and $\mu$-like
event rates including the angular information. It contains 40
data points.
\item{UP--$\mu$} Muon fluxes for stopping 
and through--going muons at Super--Kamiokande and  MACRO 
which correspond to 25 data points. 
\item{SK} The angular distribution of  $e$-like and $\mu$-event rates 
and upgoing muon fluxes measured at Super--Kamiokande. It contains  
35 data points.
\item{ALL--ATM} The full data sample of atmospheric neutrino data 
which corresponds to 65 points.
\end{itemize}

The first result of our analysis refers to the no-oscillation
hypothesis. As can be seen from the fifth column of
Table~\ref{table:chiatm}, the $\chi^2$ values in the absence of new
physics -- as obtained with our prescriptions for different
combinations of atmospheric data sets -- clearly show that the data
are totally inconsistent with the SM hypothesis.  In fact, the global
analysis, which refers to the full combination ALL--ATM, gives
$\chi^2_{SM,\text{ ALL--ATM}}=191.7/\text{(65 d.o.f.)}$ corresponding to a
probability $\lesssim 10^{-14}$. This result is rather insensitive to
the inclusion of the CHOOZ reactor data, as can be seen by comparing
the values of $\chi^2$ given in the last two lines of
Table~\ref{table:chiatm}.  This indicates that the Standard Model can
be safely ruled out.  In contrast, the $\chi^2$ for the global
analysis decreases to $\chi^2_{min}= 61.7/\text{(62 d.o.f.)}$
[$\chi^2_{min}= 62.5/\text{(63 d.o.f.)}$ including CHOOZ], acceptable at the
51\% when oscillations are assumed.

\begin{figure} 
\begin{center} 
\mbox{\epsfig{file=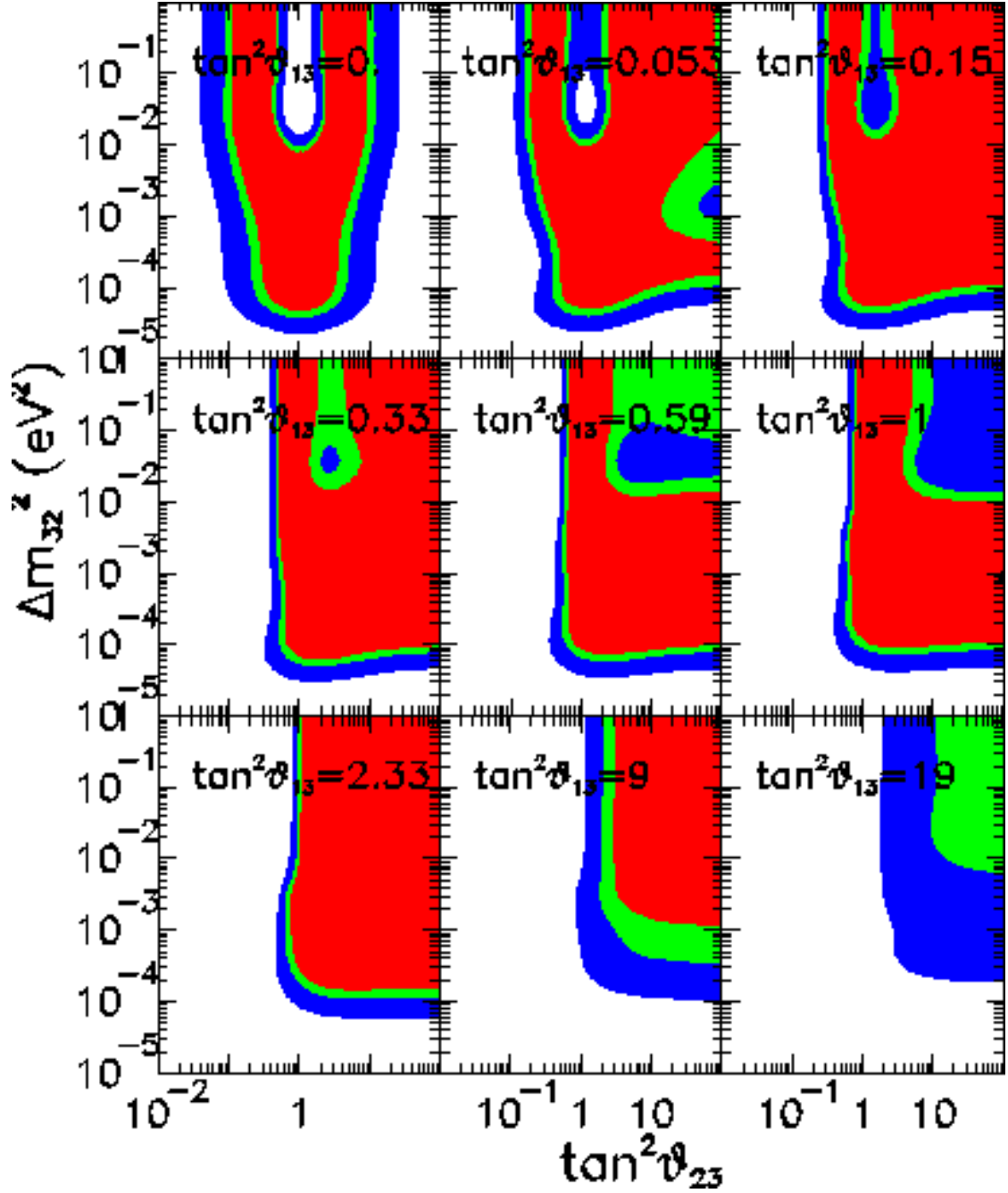,width=0.9\textwidth}} 
\end{center} 
    \vspace{3mm}
    \caption{
      Allowed $(\tan^2\theta_{23}, \Delta m^2_{32})$ regions for
      different $\tan^2\theta_{13}$ values, for the FINKS combination
      of atmospheric neutrino data.  The regions refer to 90, 95
      and 99\% CL.  The best--fit point in the three parameter space
      is denoted as a star.}
\label{fig:finks}
\end{figure}

\begin{figure} 
\begin{center} 
\mbox{\epsfig{file=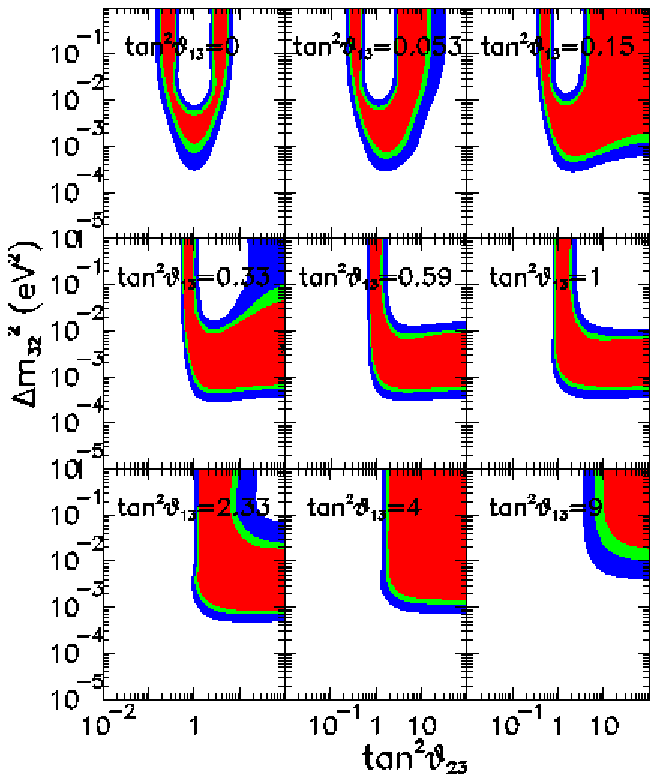,width=0.9\textwidth}} 
\end{center} 
    \vspace{3mm}
    \caption{
      Allowed three--neutrino $(\tan^2\theta_{23}, \Delta m^2_{32})$
      regions for different $\tan^2\theta_{13}$ values, for the
      CONT--UNBIN combination of atmospheric neutrino data.  The
      regions refer to 90, 95 and 99\% CL.  The best--fit point is
      denoted as a star.}
\label{fig:unbin}
\end{figure}

\begin{figure}
\begin{center} 
\mbox{\epsfig{file=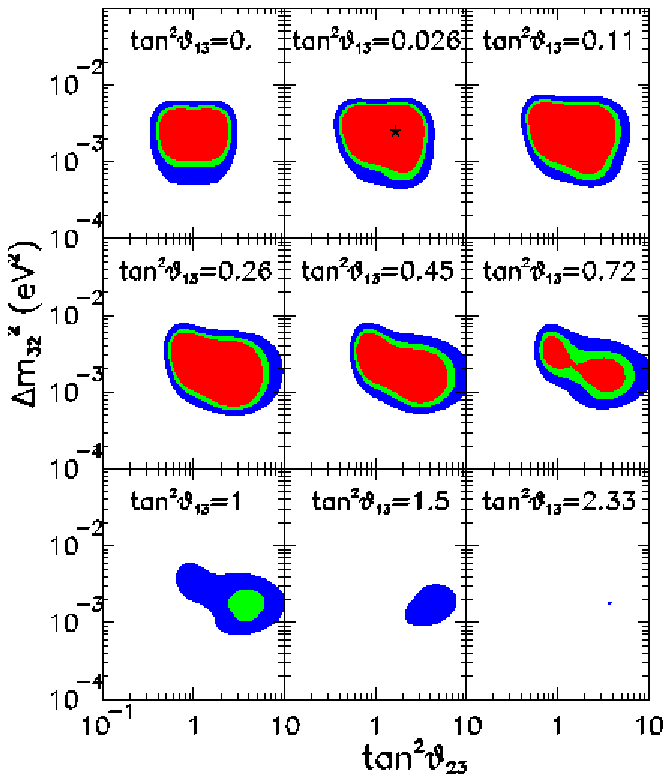,width=0.9\textwidth}} 
\end{center} 
    \vspace{3mm}
    \caption{
      Allowed three--neutrino $(\tan^2\theta_{23}, \Delta m^2_{32})$
      regions for different $\tan^2\theta_{13}$ values, for the
      CONT--BIN combination of atmospheric neutrino events The regions
      refer to 90, 95 and 99\% CL.  The best--fit point is denoted as
      a star.}
\label{fig:cont}
\end{figure}

\begin{figure} 
\begin{center}
\mbox{\epsfig{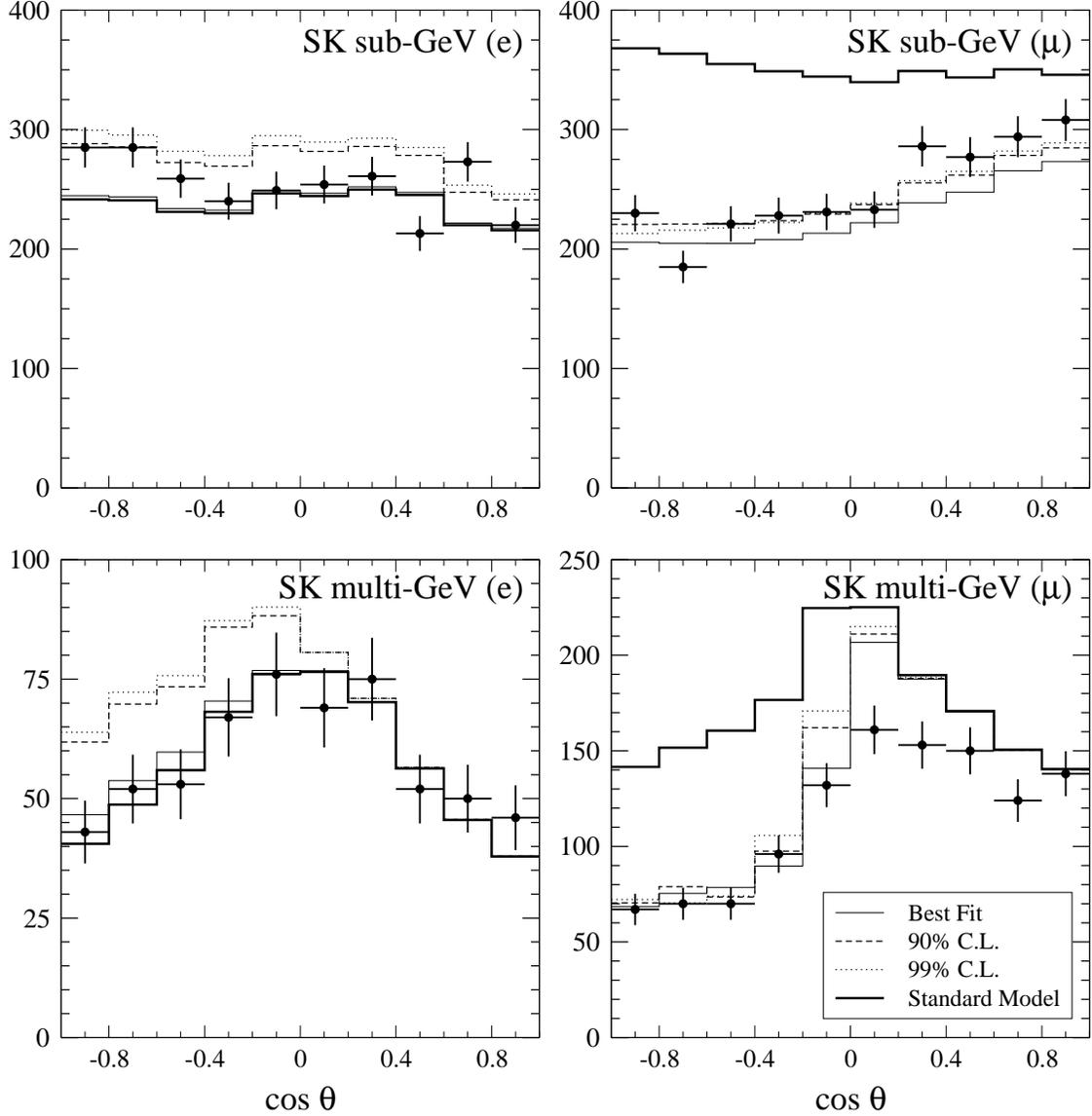}} 
\end{center}
    \vspace{3mm}
\caption{
    Zenith-angle distributions for the Super--Kamiokande $e$--like
  (left panels) and $\mu$--like (right-panels) events, both in the
  sub-GeV (upper panels) and multi-GeV (lower panels) energy range.
  The thick solid line is the expected distribution in the SM.  The
  thin full line is the prediction for the overall best-fit point of
  ALL--ATM data $\tan^2\theta_{13}=0.025$, $\Delta
  m^2_{32}=3.3\times 10^{-3}$ eV$^2$ and $\tan^2\theta_{23}=1.6$.  The
  dashed (dotted) histogram correspond to the distributions for
  $\Delta m^2_{32}=3.3\, (2.85)\times 10^{-3}$ eV$^2$,
  $\tan^2\theta_{23}=3.0\,(3.1)$ and $\tan^2\theta_{13}=0.33\, (0.54)$ which are
  allowed at 90 (99)\% CL.}

\label{fig:angcont}
\end{figure}

\begin{figure}
\begin{center} 
\mbox{\epsfig{file=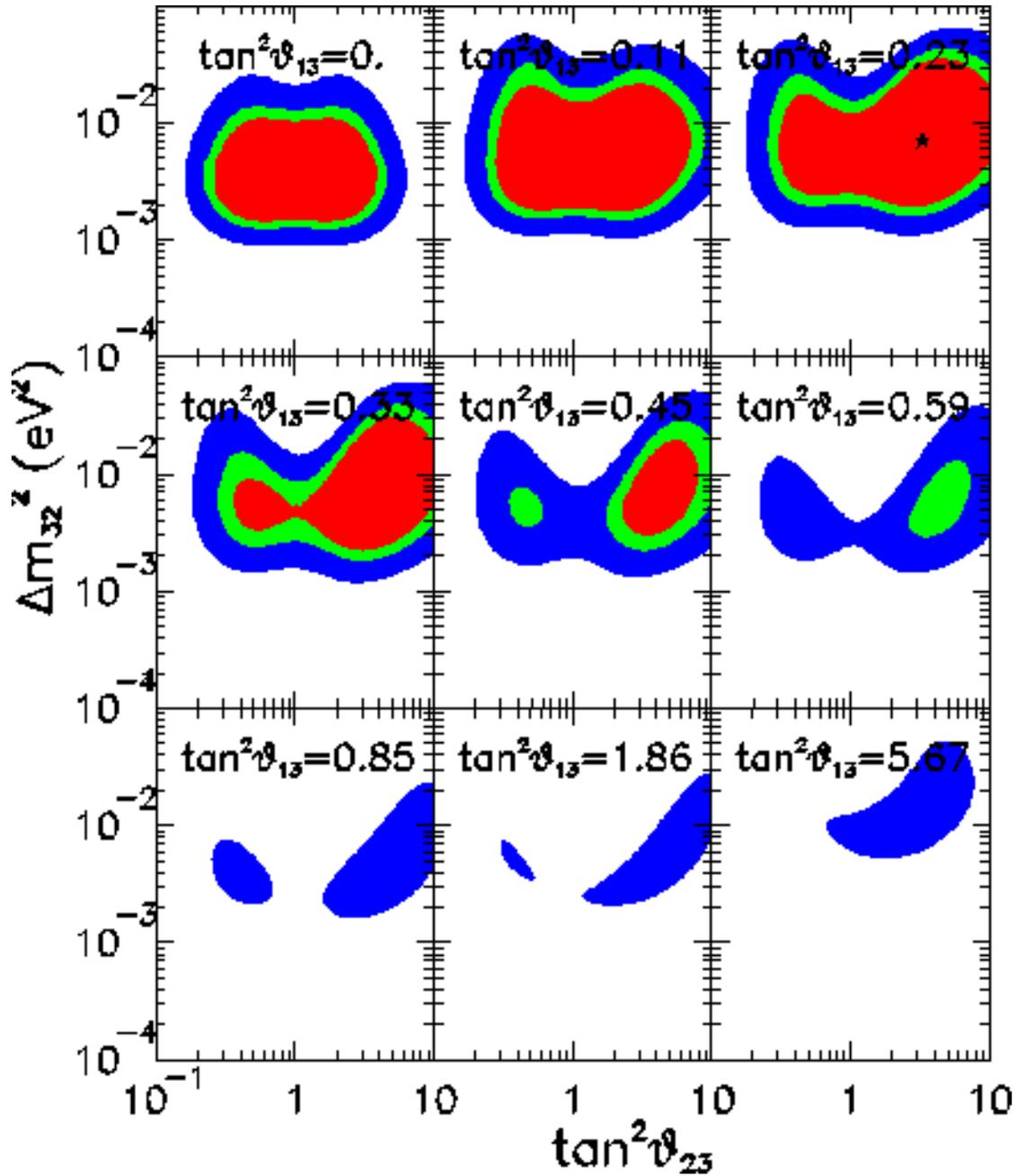,width=0.9\textwidth}} 
\end{center} 
    \vspace{3mm}
    \caption{90, 95 and 99\% 
      three--neutrino allowed regions in the $(\tan^2\theta_{23},
      \Delta m^2_{32})$ plane for different $\tan^2\theta_{13}$
      values, for the combination of UP--$\mu$ events induced by
      atmospheric neutrinos from Super--Kamiokande (through--going and
      stopping) and MACRO (through--going only).  The best--fit point is
      denoted as a star.}
    \label{fig:up}
\end{figure}

\begin{figure}
\begin{center}
\mbox{\epsfig{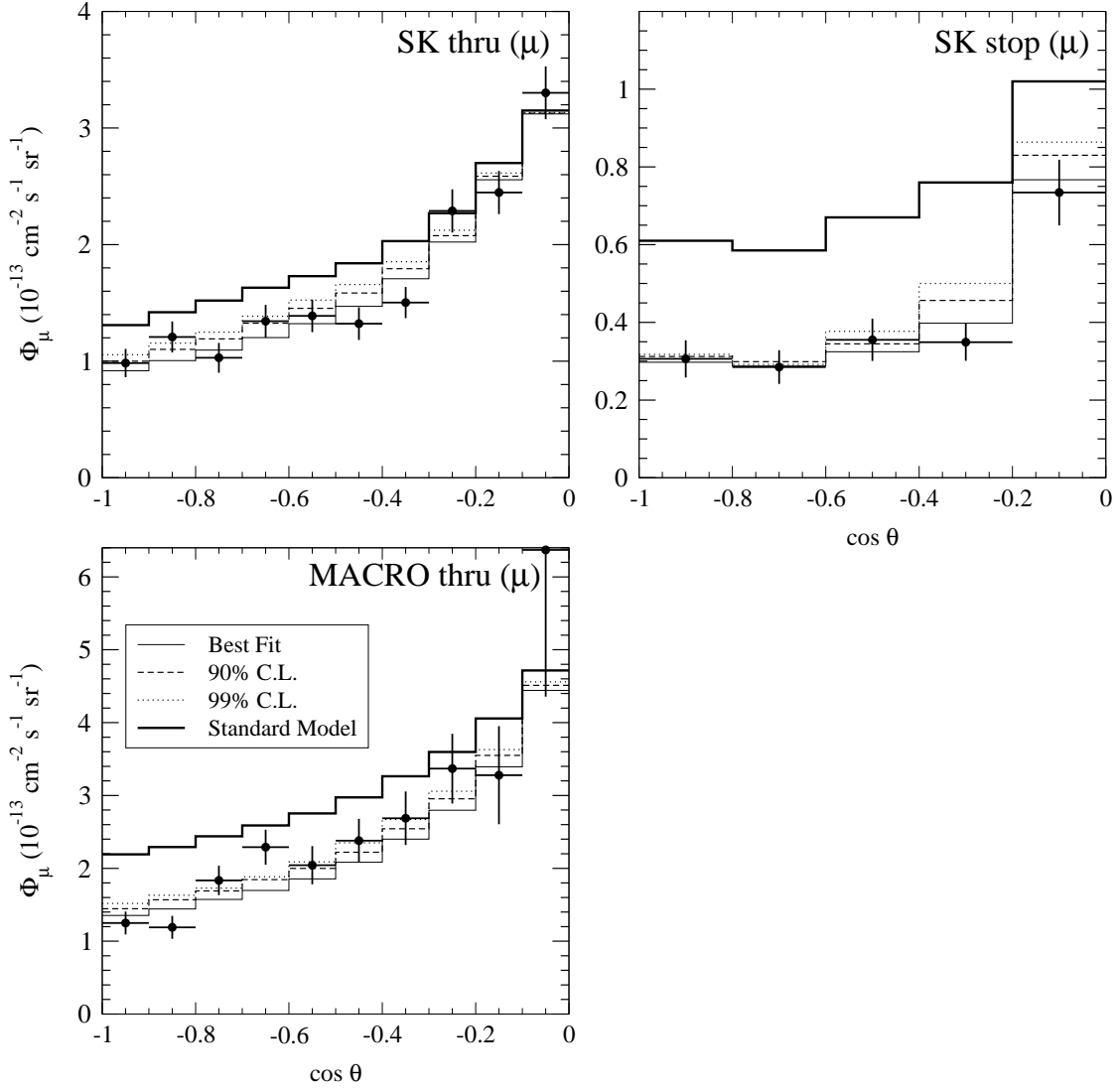}} 
\end{center}
 \vspace{3mm}
    \caption{
      Zenith-angle distributions for upward-going muon events in
      Super--Kamiokande and MACRO.
The thick solid line is the expected distribution in the SM. 
The thin full line is the prediction for the overall best-fit point of 
ALL--ATM data $\tan^2\theta_{13}=0.025$, 
$\Delta m^2_{32}=3.3\times 10^{-3}$ eV$^2$ and $\tan^2\theta_{23}=1.6$.
The dashed (dotted) histogram correspond to the distributions for 
$\Delta m^2_{32}=3.3\, (2.85)\times 10^{-3}$ eV$^2$, 
$\tan^2\theta_{23}=3.0\,(3.1)$ and $\tan^2\theta_{13}=0.33\, (0.54)$ which
are allowed at 90 (99)\% CL.}
    \label{fig:angup}
\end{figure}

\begin{figure}
\begin{center} 
\mbox{\epsfig{file=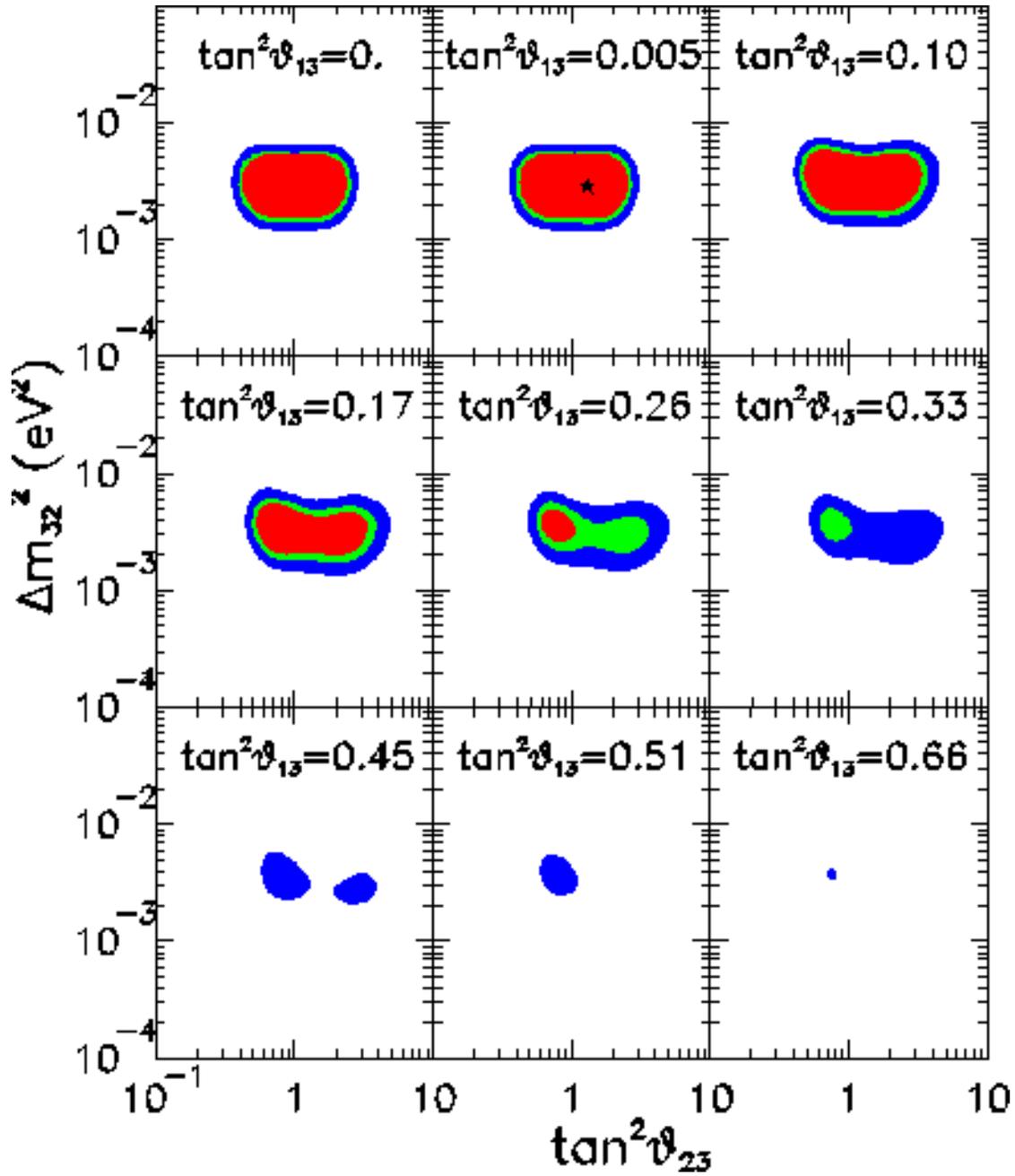,width=0.9\textwidth}} 
\end{center} 
    \vspace{3mm}
    \caption{
      Allowed $(\tan^2\theta_{23}, \Delta m^2_{32})$ regions for
      different $\tan^2\theta_{13}$ values, for the combination of SK
      atmospheric neutrino events.  The regions refer to 90, 95 and
      99\% CL.  The best--fit point is denoted as a star.}
\label{fig:sk}
\end{figure}

\begin{figure}
\begin{center} 
\mbox{\epsfig{file=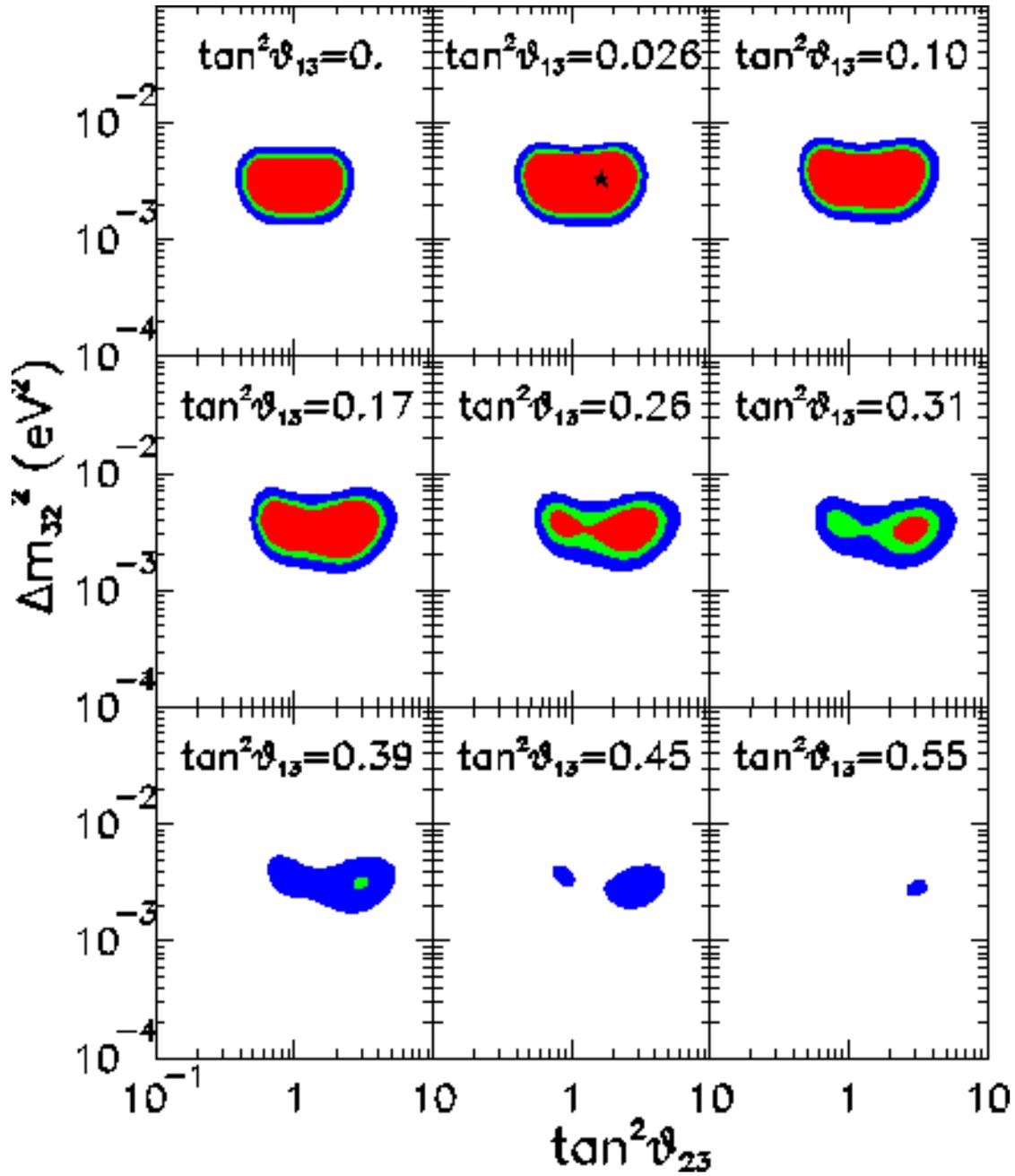,width=0.9\textwidth}} 
\end{center} 
    \vspace{3mm}
    \caption{90, 95 and 99\% CL
      three--neutrino allowed regions in $(\tan^2\theta_{23}, ~\Delta
      m^2_{32})$ for different $\tan^2\theta_{13}$ values, for the
      combination of ALL--ATM neutrino data.  The best--fit point is
      denoted as a star.}
    \label{fig:all}
\end{figure}

Table~\ref{table:chiatm} also gives the minimum $\chi^2$ values and
the resulting best fit points for the various combinations of data
sets considered.  Note that for FINKS, CONT--UNBIN and UP--$\mu$
combinations the best-fit point is characterized by a rather large
value of $\tan^2\theta_{13}$, while all the other data set favour a
value very close to 0. The corresponding allowed regions for the
$(\tan^2\theta_{23}, ~\Delta m^2_{32})$ oscillation parameters at 90,
95 and 99\% CL for the different combinations are depicted in the set
of Figures from Fig.~\ref{fig:finks}--Fig.~\ref{fig:all}. In all these
figures the upper-left panel, $\tan^2\theta_{13} = 0$, corresponds to
pure $\nu_\mu \to \nu_\tau$ oscillations, and one can note the exact
symmetry of the contour regions under the transformation $\theta_{23}
\to \pi/4 - \theta_{23}$. This symmetry follows from the fact that in
the pure $\nu_\mu \to \nu_\tau$ channel matter effects cancel out and
the oscillation probability depends on $\theta_{23}$ only through the
double--valued function $\sin^2(2\theta_{23})$.  For non-vanishing
values of $\theta_{13}$ this symmetry breaks due to the
three--neutrino mixing structure even if matter effects are neglected.
With our sign assignment we find that in most cases for non-zero values
of $\theta_{13}$ the allowed regions become larger in the second
octant of $\theta_{23}$.

In Figs.~\ref{fig:finks} and~\ref{fig:unbin} we present the allowed
regions in ~$(\tan^2\theta_{23}, ~\Delta m^2_{32})$ for different
values of $\tan^2\theta_{13}$, for the FINKS and CONT--UNBIN data
respectively. It is evident that, despite of the large statistics
provided by Super--Kamiokande data, it is not possible from the
information on the total event rates only, without including the
angular dependence, to place any upper bound on $\Delta m^2_{32}$, and
even the lower bound $\Delta m^2_{32} > 2 \times 10^{-4}$~eV$^2$ is
rather weak.  The CONT--UNBIN data also do not provide a relevant
constraint on $\tan^2\theta_{13}$.

Conversely, in Fig.~\ref{fig:cont} we display the allowed regions in
$(\tan^2\theta_{23}, ~\Delta m^2_{32})$ for different values of
$\sin^2\theta_{13}$, for the combination of CONT--BIN events,
including also the information on the angular distributions.  Note
that, as expected, the {\it upper} bound on $\Delta m^2_{32}$ is now
rather strong (better than $10^{-2}$~eV$^2$) as a consequence of the
fact that no suppression for downgoing $\nu_\mu$ neutrinos is
observed. This imposes a lower bound on the neutrino oscillation
length and consequently an upper bound on the mass difference. However
contained events alone still allow values of $\Delta m^2_{32} <
10^{-3}$ eV$^2$.  Note also that the allowed region is still rather
large for $\tan^2\theta_{13} \approx 0.7$ and at 99\% CL it only
disappears for $\tan^2\theta_{13} \simeq 2.4$.

In order to illustrate the main effect of adding the angle
$\theta_{13}$ in the description of the angular distribution of
contained atmospheric neutrino events we show in
Fig.~\ref{fig:angcont} the zenith-angle distributions for the
Super--Kamiokande $e$--like (left panels) and $\mu$--like 
(right-panels) contained events, both in the sub-GeV (upper panels)
and multi-GeV (lower panels) energy range.  The thick solid line is
the expected distribution in the absence of oscillation (SM
hypothesis), while the thin full line represents the prediction for
the overall best-fit point of the full atmospheric data set (ALL--ATM)
(see also Table~\ref{table:chiatm}) which occurs at a small
$\tan^2\theta_{13}=0.026$ (with $\Delta m^2_{32}=3.3\times 10^{-3}$
eV$^2$, $\tan^2\theta_{23}=1.63$).  The dashed and dotted
histograms correspond to the distributions with increasing value of
$\tan^2\theta_{13}=0.33,0.54$ which are the maximum allowed values at
90 and 99\% CL from the analysis of all atmospheric data.  For each
such $\tan^2\theta_{13}$ we choose $\Delta m^2_{32} $ and
$\tan^2\theta_{32}$ so as to minimize the $\chi^2$.
Clearly the oscillation description is excellent as long as the
oscillation is mainly in the $\nu_\mu \to \nu_\tau$ channel (small
$\theta_{13}$).  This is simply understood since, from the left
panels, it is clear that $e$-like events are well accounted for within
the no-oscillation hypothesis. From the figure we see that increasing
$\tan^2\theta_{13}$ leads to an increase in all the contained event
rates.  This is due to the fact that an increasing fraction of
$\nu_\mu$ now oscillates as $\nu_\mu\to\nu_e$ (also $\nu_e$'s
oscillate as $\nu_e\to\nu_\mu$ but since the $\nu_e$ fluxes are
smaller this effect is relatively less important) spoiling the good
description of the $e$--type data, especially for upgoing
multi-GeV electron events. For multi-GeV events all the curves
coincide with the SM one for downgoing neutrinos which did not have
the time to oscillate. For sub-GeV this effect is lost due to the
large opening angle between the neutrino and the detected lepton. We
also see that for multi-GeV electron neutrinos the effect of
$\theta_{13}$ is larger close to the vertical ($\cos\theta=-1$) where
the expected ratio of fluxes in the SM $R(\nu_\mu/\nu_e)$ is larger.
Conversely the relative effect of $\theta_{13}$ for $\nu_\mu$ is
larger close to the horizontal direction, $\cos\theta=0$.

Now we move to upgoing muon events. In Fig.~\ref{fig:up} we show the
allowed regions in $(\tan^2\theta_{23}, ~\Delta m^2_{32})$ for
different $\tan^2\theta_{13}$ values, for the combination UP-$\mu$
which contains upgoing-muon events from Super--Kamiokande (through--going
and stopping) and MACRO (through--going only).  This plot is complementary
to Fig.~\ref{fig:cont} (corresponding to the CONT--BIN combination),
in the sense that the data in CONT--BIN and UP-$\mu$ combinations are
completely disjoint. In contrast to the CONT--BIN case, now the upper
bound on $\Delta m^2_{32}$ is much weaker ($\approx 3 \times
10^{-2}$~eV$^2$), while the lower bound is now stronger. Again, no
relevant bound can be put on $\theta_{13}$ from the analysis of
upgoing events alone.

The angular distribution for the upward-going muon fluxes for
increasing values of $\theta_{13}$ is presented in
Fig.~\ref{fig:angup}.  The thick solid line is the expected
distribution in the absence of oscillations (SM hypothesis), while the
thin full line represents the prediction for the overall best-fit
point of all atmospheric data (ALL--ATM).  As in Fig.~\ref{fig:angcont}
the dashed and dotted histograms correspond to the distributions with
increasing value of $\tan^2\theta_{13}=0.33,0.54$ (maximum acceptable
values at 90 and 99\% CL from the analysis of ALL--ATM data).  From
the figure we see that the effect of adding a large $\theta_{13}$ in
the expected upward muon fluxes is not very significant. For stopping
muons the effect is larger for neutrinos arriving horizontally.  This
is due, as for the case of multi-GeV muons, to the larger
$R(\nu_e/\nu_\mu)$ SM flux ratio in this direction which implies a
larger relative contribution from $\nu_e$ oscillating to $\nu_\mu$.
This feature is lost in the case of through--going muons because this
effect is partly compensated by the matter effects and also by the
increase of $\tan^2\theta_{23}$.

In order to perform a separate critical analysis of the implications
of all Super--Kamiokande data by themselves we show in Fig.~\ref{fig:sk}
the allowed regions in $(\tan^2\theta_{23}, ~\Delta m^2_{32})$ for
different $\tan^2\theta_{13}$ values, for the combination of SK data
(contained and upgoing). Due to the large statistics provided by this
experiment, $\Delta m^2_{32}$ is strongly bounded both from above and
from below. Moreover no region of parameter space is allowed (even at
99\% CL)  for $\tan^2\theta_{13} \gtrsim 0.7$.  It is also
interesting to notice that (unlike in the ALL--ATM combination
discussed later) as $\tan^2\theta_{13}$ increases the allowed region
in the second octant of $\theta_{23}$ becomes smaller and finally
disappears. This behaviour is driven by the SK contained event data
which favours the first $\theta_{23}$ octant as can be seen in the
corresponding line in Table \ref{table:chiatm}.

\begin{figure}
\begin{center} 
\mbox{\epsfig{file=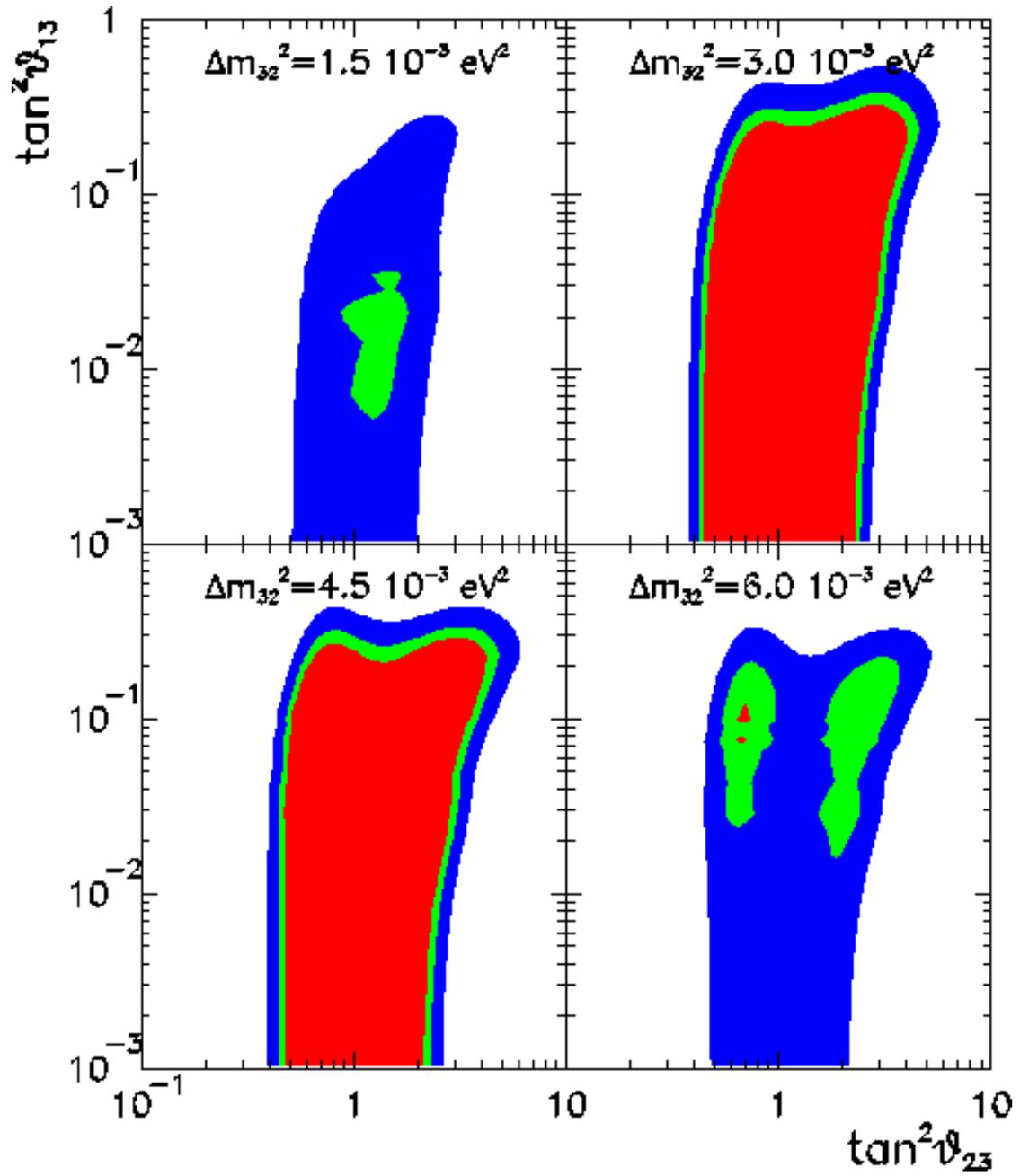,width=0.9\textwidth}} 
\end{center} 
    \vspace{3mm}
    \caption{
      Allowed $(\tan^2\theta_{23}, ~\tan^2\theta_{13})$ regions for
      different $\Delta m^2_{32}$ values, for the combination 
      ALL--ATM of  atmospheric neutrino data.}
    \label{fig:ttall}
\end{figure}

Finally we discuss the results from the combined global analysis of
all the atmospheric neutrino data (ALL--ATM). The allowed range of
parameters are shown in Figs.~\ref{fig:all} and~\ref{fig:ttall}. In
Figs.~\ref{fig:all} we show the global $(\tan^2\theta_{23}, ~\Delta
m^2_{32})$ allowed regions, for different values of
$\tan^2\theta_{13}$ while in Fig.~\ref{fig:ttall} we show the
corresponding projection of the three--dimensional parameter space in
the $(\tan^2\theta_{23},~\tan^2\theta_{13})$ plane, for different
values of $\Delta m^2_{32}$.  Although Fig.~\ref{fig:all} shows no
qualitative difference with respect to the allowed regions from the
analysis of SK data alone displayed in Fig.~\ref{fig:sk}, we find that
the inclusion of the other experimental results still results into an
slightly tighter restriction on the allowed parameter space. In this
way, for instance, we find that the allowed region (at 99\% CL)
disappears for $\tan^2\theta_{13} \simeq 0.6$.  Moreover, as mentioned
before comparing Fig.~\ref{fig:sk} with \ref{fig:all} we see a slight trend
in Fig.~\ref{fig:sk} towards the $\tan^2\theta_{23} <1$ octant, while
$\tan^2\theta_{23} > 1$ is favoured in the other case.

\begin{figure} 
\begin{center} 
\mbox{\epsfig{file=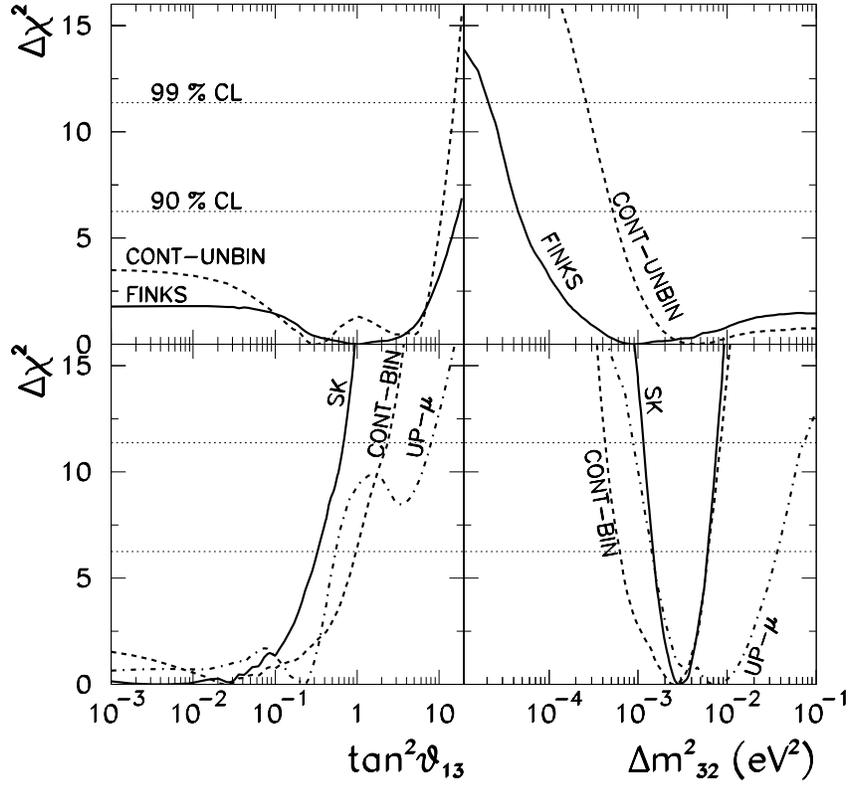,width=0.7\textwidth}} 
\end{center} 
    \vspace{3mm}
    \caption{
      Dependence of the $\Delta \chi^2$ function on
      $\tan^2\theta_{13}$ and on $\Delta m^2_{32}$ , for different
      combinations of atmospheric neutrino events.  The dotted
      horizontal lines correspond to the 90\%, 99\% CL limits.}
    \label{fig:chi_atm}
\end{figure}

\begin{figure} 
\begin{center} 
\mbox{\epsfig{file=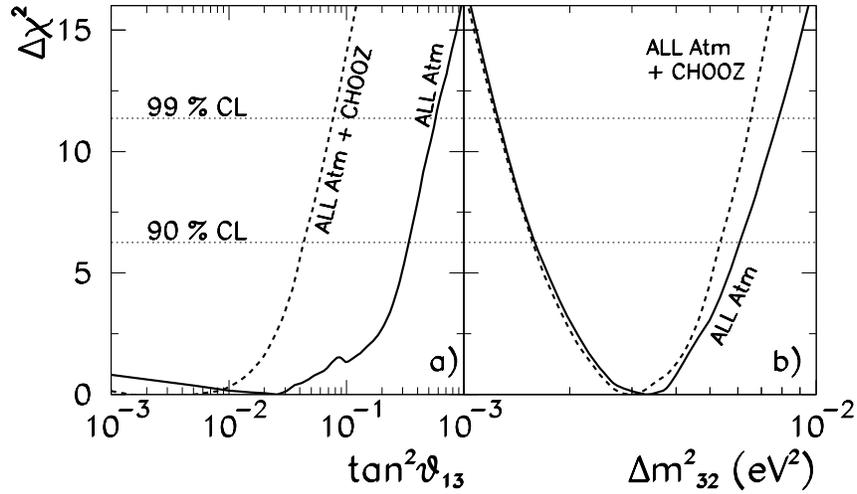,width=0.7\textwidth}} 
\end{center} 
    \vspace{3mm}
    \caption{
      Dependence of the $\Delta \chi^2$ function on
      $\tan^2\theta_{13}$ and on $\Delta m^2_{32}$, for the
      combination ALL--ATM of atmospheric neutrino data.  The dashed
      line includes also the data from CHOOZ.  The dotted horizontal
      lines correspond to the 90\%, 99\% CL limits for 3~d.o.f.}
    \label{fig:chi_atmall}
\end{figure}

All these features can be more quantitatively observed in
Figs.~\ref{fig:chi_atm} and~\ref{fig:chi_atmall} where we show the
dependence of the $\Delta\chi^2_{atm}$ function on $\tan^2\theta_{13}$
and on $\Delta m^2_{32}$, for the
different combination of atmospheric neutrino events.  In these plots
all the neutrino oscillation parameters which are not displayed have
been ``integrated out'', or, equivalently, the displayed
$\Delta\chi^2$ is minimized with respect to all the non-displayed
variables. From the left panels of Figs.~\ref{fig:chi_atm}
and~\ref{fig:chi_atmall} we can read the upper bound on $\tan^2\theta_{13}$
that can be extracted from the analysis of the different samples of
atmospheric data alone regardless of the values of the other
parameters of the three--neutrino mixing matrix. The corresponding 90
and 99\% CL bounds are listed in Table~\ref{table:limits}. Conversely
from the right panels of Figs.~\ref{fig:chi_atm}
and~\ref{fig:chi_atmall} we extract the allowed value of $\Delta
m^2_{32}$ by the different combination irrespective of the of the
values of the other parameters of the three--neutrino mixing matrix.

In conclusion we see that the analysis of the full atmospheric neutrino 
data in the framework of three--neutrino oscillations clearly favours
the $\nu_\mu \to \nu_\tau$ oscillation hypothesis. As a matter of fact 
the best fit corresponds to a small value of $\theta_{13}= 9^\circ$.
But it still allows for a non-negligible $\nu_\mu \to \nu_e$ 
component. More quantitatively we find that the following ranges of 
parameters are allowed at 90 [99]~\% CL from this analysis
\begin{eqnarray}
[1.25\times 10^{-3}]\, 1.6 \times 10^{-3}\;
& <\; \Delta m^2_{32}/\mbox{\rm eV$^2$}\;< \;& 
\;6\times 10^{-3}\, [8\times 10^{-3}] \\ \nonumber
[0.37]\, 0.43\;&<\;\tan^2\theta_{23}\;<&\; 4.2 \, [6.2] \\\label{atmranges}
&\;\;\tan^2\theta_{13} \;<\;&  0.34 \,[0.57] \nonumber
\end{eqnarray}
One must take into account that these ranges are strongly correlated
as illustrated in Figs.~\ref{fig:all} and~\ref{fig:ttall}.

\subsection{Fit to Atmospheric and CHOOZ data}

\begin{figure}
\begin{center} 
\mbox{\epsfig{file=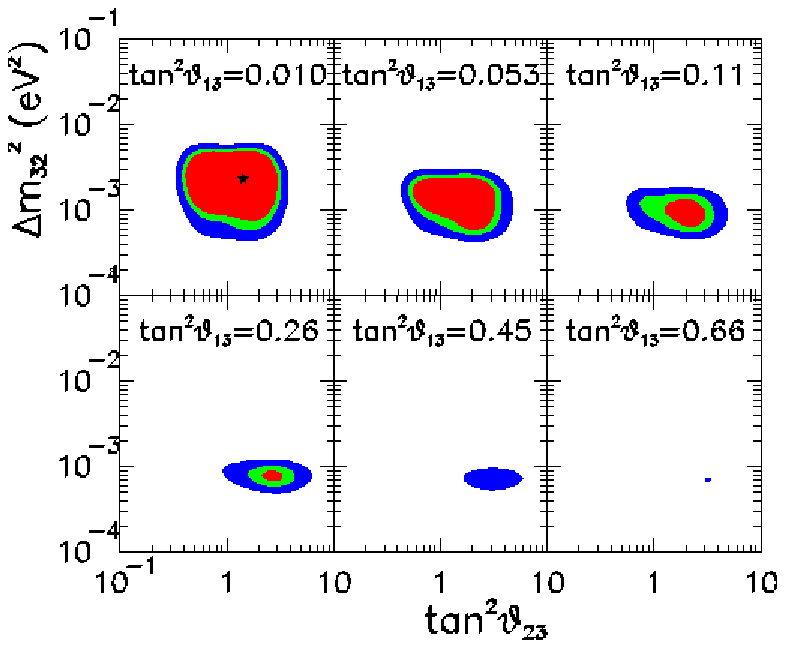,width=0.9\textwidth}} 
\end{center} 
    \vspace{3mm}
    \caption{
      Same as Fig.~{\protect{\ref{fig:cont}}}, but including also the
      CHOOZ result. The best--fit point is denoted as a star.}
    \label{fig:cont.CH}
\end{figure}

\begin{figure}
\begin{center} 
\mbox{\epsfig{file=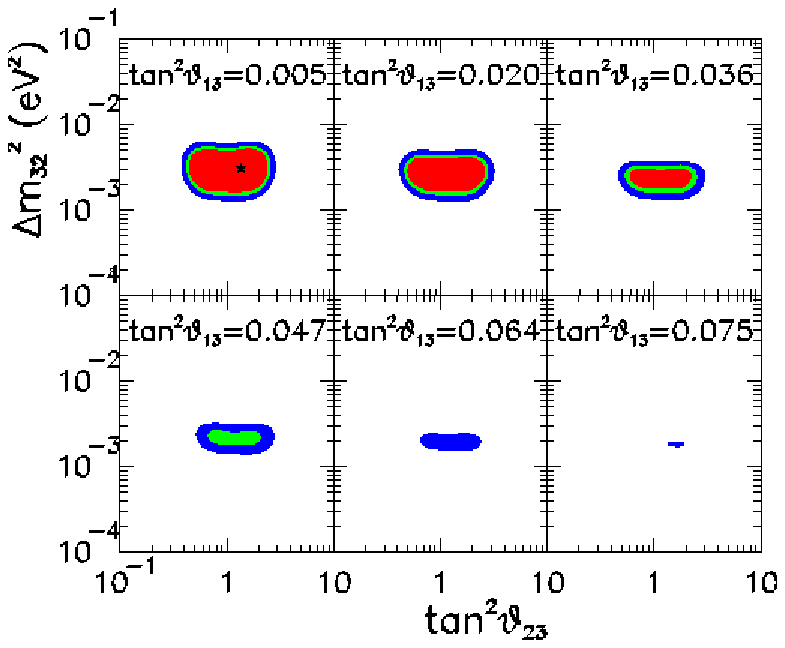,width=0.9\textwidth}} 
\end{center} 
    \vspace{3mm}
    \caption{
      Same as Fig.~{\protect{\ref{fig:all}}}, but including also the
      CHOOZ result. The best--fit point is denoted as a star.}
    \label{fig:all.CH}
\end{figure}

\begin{figure}
\begin{center} 
\mbox{\epsfig{file=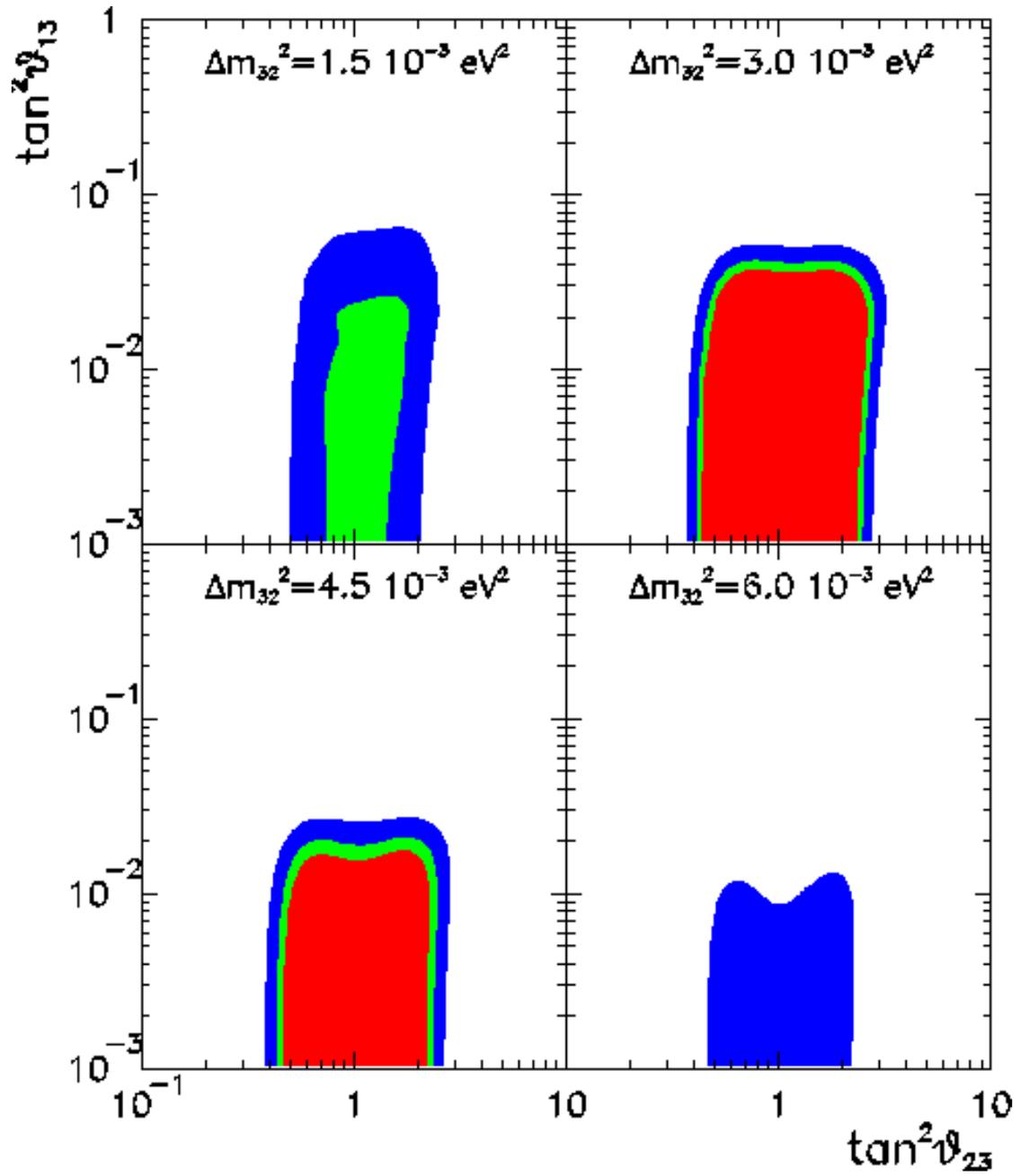,width=0.9\textwidth}} 
\end{center} 
    \vspace{3mm}
    \caption{
      Same as Fig.~{\protect{\ref{fig:ttall}}}, but including also the 
      CHOOZ result.}
    \label{fig:ttall.CH}
\end{figure}

We now describe the effect of including the CHOOZ reactor data
together with the atmospheric data samples in a combined 3--neutrino
$\chi^2$ analysis. The results of this analysis are summarized in
Fig.~\ref{fig:chi_atmall},
Figs.~\ref{fig:cont.CH}--\ref{fig:ttall.CH}, and in
Tables.~\ref{table:limits} and~\ref{table:chiatm}. In this analysis
we will assume that  $\Delta m^2_{21}\lesssim 3\times 10^{-4}$ eV$^2$
and work under the approximations in
Eqs.~(\ref{eq:evolap.1}) and~(\ref{eq:evolap.2}).

As discussed in Sec.~\ref{data:CHOOZ} the negative results of the
CHOOZ reactor experiment strongly disfavour the region of parameters
with $\Delta m^2_{32}\gtrsim 10^{-3} $ eV$^2$ and
$\sin^2(2\theta_{13}) \gtrsim 0.10$ ($0.026 \lsim
\tan^2\theta_{13} \lsim 38$).  However for smaller values of
$\Delta m^2_{32}$ the CHOOZ result leads to much weaker bounds on the
$\theta_{13}$ mixing angle.  To illustrate this point we show in
Fig.~\ref{fig:cont.CH}, the allowed regions from the combination of
the CONT--BIN events with the CHOOZ data. We see in
Fig.~\ref{fig:cont.CH}, which should be compared with
Fig.~\ref{fig:cont}, that, as soon as $\tan^2\theta_{13}$ deviates from
zero the $\Delta m^2_{32} > 10^{-3}$~eV$^2$ region is ruled out.
However there is still a region in the parameter space which survives
(at 99\% CL) up to $\tan^2\theta_{13} \approx 0.66$. Thus, even with
the large statistics provided by the Super--Kamiokande data and
including the CHOOZ result, it is not possible to constrain
$\theta_{13}$ using only the data on contained events.

The situation is changed once the upgoing muon events are included in
the analysis. As shown in Figs.~\ref{fig:up} and~Fig.~\ref{fig:all},
the upgoing muon data disfavours the low mass $\Delta
m^2_{32}<10^{-3}$ eV$^2$.  As a result the full 99\% CL allowed
parameter regions from the global analysis of the atmospheric data
shown in Fig.~\ref{fig:all} lies in the mass range where the CHOOZ
experiment should have observed oscillations for sizeable
$\theta_{13}$ values.  This results into the shift of the global
minimum from the combined atmospheric plus CHOOZ data to
$\theta_{13}=0$.  Thus adding the reactor data (Fig.~\ref{fig:all.CH}) has
as main effect effect the strong improvement of the
$\tan^2\theta_{13}$ limit, as seen from the left panel in
Fig.~\ref{fig:chi_atmall} and by comparing the allowed ranges in
Fig.~\ref{fig:ttall} and Fig.~\ref{fig:ttall.CH}.  From these figures
we see that the higher $\Delta m^2_{32}$ the more the CHOOZ data
restricts $\theta_{13}$. In this way the allowed range of
$\theta_{13}$ at 99\% CL is reduced by a factor $\sim$ 4, 10, 15, 25
for $\Delta m^2_{32}=1.5,3.0,4.5$ and 6$\times 10^{-3}$ eV$^2$
respectively by the inclusion of the CHOOZ data.  In contrast, the
allowed range of $\Delta m^2_{32}$ is only mildly restricted when
combining the full atmospheric data with the reactor result as
displayed in the right panel of Fig.~\ref{fig:chi_atmall}.

We get as final results from the atmospheric and reactor neutrino data 
analysis the following allowed ranges of parameters at 90 [99]~\% CL 
\begin{eqnarray}
[1.2\times 10^{-3}]\, 1.6 \times 10^{-3}\;
<& \Delta m^2_{32}/\mbox{\rm eV$^2$}&< \; 
\;5.4\times 10^{-3}\, [6.6\times 10^{-3}] \\ \nonumber
[0.36]\, 0.43<&\tan^2\theta_{23}&<\; 2.7 \, [3.3] \\\label{atmreranges}
&\tan^2\theta_{13} &<\;  0.043 \,[0.08] \nonumber
\end{eqnarray}
Comparing with the allowed ranges shown in Eq.~(\ref{atmranges}) we
see that the main effect of the addition of the CHOOZ results to the
atmospheric neutrino data analysis is the stronger constraint on the
allowed value of the angle $\theta_{13}$. This illustrates again the
fact that the full allowed region of $\Delta m^2_{32}$ from the
atmospheric neutrino analysis lies in the sensitivity range for the
CHOOZ reactor.  However it only holds once upgoing muons are
included in the fit.
Including the CHOOZ results has very little effect on the allowed
range of $\Delta m^2_{32}$ while it results into a tighter constraint
on $\tan^2\theta_{23}$ on the second octant. This last effect arises
from the fact that in order to have $\tan^2\theta_{23}$ far into the
second octant one requires large $\theta_{13}$ values which are
forbidden by the CHOOZ data.

\section{Combined Solar, Atmospheric and Reactor Analysis}
\label{analysis:combined}

\begin{figure} 
\begin{center} 
\mbox{\epsfig{file=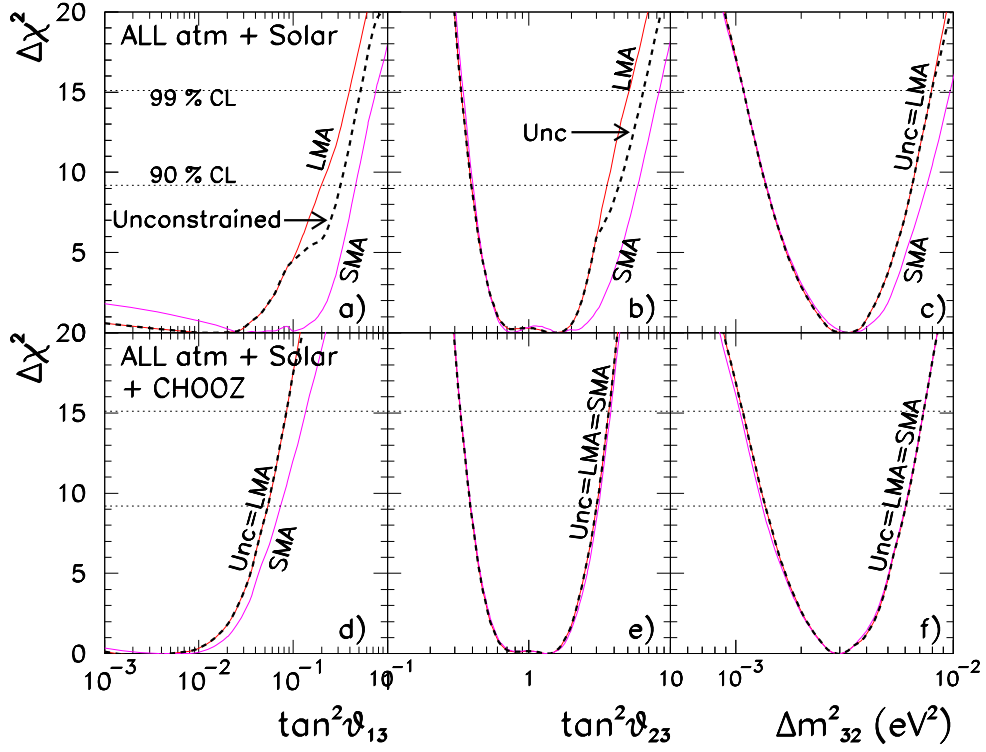,width=0.8\textwidth}} 
\end{center} 
\caption{Dependence of $\Delta \chi^2$ on $\tan^2\theta_{13}$,
  $\tan^2\theta_{23}$ and on $\Delta m^2_{32}$ , for the analysis of
  atmospheric and solar neutrino events (upper panels) and
  atmospheric, solar and CHOOZ data (lower panels).  The dotted
  horizontal lines correspond to the 90\% and 99\% CL limits for
  5~d.o.f.}
\label{fig:chiglo} 
\end{figure}

\begin{figure} 
\begin{center} 
\mbox{\epsfig{file=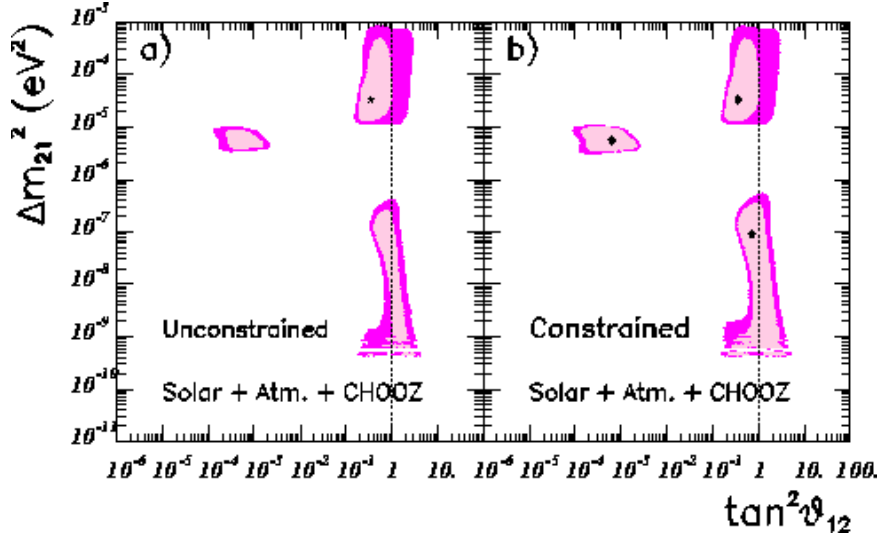,width=0.7\textwidth}} 
\end{center} 
\caption{Allowed regions in  $\Delta{m}^2_{21}$ and $\tan^2\theta_{12}$  
  from the global analysis of solar, atmospheric and reactor neutrino
  data.  (a) Regions for the unconstrained analysis defined in terms
  of the increases of $\Delta\chi^2$ for 5~d.o.f.\ from the global
  best fit point denoted as a star. (b) Regions for the constrained
  analysis defined in terms of the increases of $\Delta\chi^2$ for
  5~d.o.f.\ from the local best fit point denoted as a dot}.
\label{fig:glosol} 
\end{figure}

In this section we describe the results of the combined analysis of
the solar and atmospheric neutrino data by themselves and also in
combination with the reactor results. In order to perform such an
analysis we have added the $\chi^2$ functions for each data set. In
this way we define
\begin{eqnarray}
\chi^2_{\text{atm + solar}}(\Delta m^2_{21},\Delta m^2_{32},\theta_{12},
\theta_{23},\theta_{13})&=& 
\chi^2_{\odot,\text{global}}(\Delta m^2_{21},\theta_{12},\theta_{13})+ 
\\\nonumber
& & \chi^2_{atm,\text{ALL}}
(\Delta m^2_{32},\theta_{23},\theta_{13}) \\\nonumber
\chi^2_{\text{atm + solar + CHOOZ}}(\Delta m^2_{21},\Delta m^2_{32},
\theta_{12},\theta_{23},\theta_{13})&=&
\chi^2_{\odot,\text{global}}(\Delta m^2_{21},\theta_{12},\theta_{13})+ 
\\\nonumber
& & 
\chi^2_{atm,\text{ALL}}(\Delta m^2_{32},\theta_{23},\theta_{13}) +
\\\nonumber
& & 
+ \chi^2_{\text{CHOOZ}}
(\Delta m^2_{21},\theta_{12},\Delta m^2_{32},\theta_{13}).
\label{eq:chicomb}
\end{eqnarray}
Notice that the $\chi^2$ functions defined in this way depend on 5
parameters. Therefore the allowed parameter space at a given CL is a
5--dimensional volume defined by the corresponding conditions on
$\Delta\chi^2$ for 5~d.o.f., $\Delta\chi^2\leq 9.24$ (11.07) [15.09] at 90
(95) [99]~\% CL. 
The results of the global combined analysis are summarized in
Fig.~\ref{fig:chiglo} and Fig.~\ref{fig:glosol} and in
Tables~\ref{table:chicomb} and ~\ref{table:limits_comb}.

In Fig.~\ref{fig:chiglo} we plot the behaviour of the $\Delta\chi^2$
functions defined above with respect to $\Delta m^2_{32}$,
$\tan^2\theta_{23}$ and $\tan^2\theta_{13}$. In the upper panels we
show the behaviour when including in the analysis only the solar and
atmospheric neutrino data while in the lower panels the CHOOZ reactor
constraint is added.  

In constructing these $\Delta\chi^2$ functions we have subtracted the
minimum in the 5--parameter space and we have minimized with respect
to the ``solar'' parameters, $\Delta m^2_{21}$ and
$\tan^2\theta_{12}$, as well as the other two undisplayed atmospheric
parameters.  Both the 5-parameter minimum and the solar minimization
have been performed either in the full $(\Delta m^2_{21},
\tan^2\theta_{12})$ parameter plane (which we previously defined as
unconstrained) or restricting the solar parameters to lie in the LMA
or SMA sectors (as defined in Eqs.~(\ref{solarrangeslma})
and~(\ref{solarrangessma}), see also the discussion below on the solar
parameter space).  The corresponding values of $\chi^2_{min}$ and the
position of the minima in the 5-dimensional space for each case are
given in Table~\ref{table:chicomb}.
We have verified that for all curves in Fig.\ref{fig:chiglo} 
the minimization in the solar parameters always occurs at 
$\Delta m^2_{21}< 10^{-4}$ eV$^2$ which ensures the validity
of the approximations in
Eqs.~(\ref{eq:evolap.1}) and~(\ref{eq:evolap.2}).

Let us first discuss the effect of the combination of solar,
atmospheric and reactor data on the common parameter to the full
analysis, $\theta_{13}$.  In Fig.~\ref{fig:chiglo}.a we display the
dependence of $\Delta\chi^2_{\text{atm + solar}}$ with
$\tan^2\theta_{13}$ once we have minimized in the other four
parameter.  As can be seen from the figure 
$\Delta\chi^2_{\text{atm + solar}}$ is sensitive to the particular 
solution of the
solar neutrino problem LMA, SMA or LOW.  As we discussed in
Sec.~\ref{analysis:solar} $\tan^2\theta_{13} \simeq 0.1$ is the
turning point where the minimum for the solar neutrino analysis
switches from LMA to SMA. This change produces the features of the
unconstrained curve. Notice that for $\tan^2\theta_{13}\gtrsim 0.1$
the unconstrained curve does not coincide with the SMA one because the
5-dimensional minimum subtracted is different. But, as expected, they
are roughly paralell. The corresponding 90 and 99\% limits can be read
in Table~\ref{table:limits_comb}.  Correspondingly
Fig.~\ref{fig:chiglo}.d shows the results once the CHOOZ data is
added. We see that the inclusion of CHOOZ produces a tighter
restriction in $\tan^2\theta_{13}$, now bounded to lie below 0.1 at 99\% CL.
As a result the unconstrained minimum is always in the LMA
region. On the other hand SMA prefers finite $\theta_{13}$ leading to
a less restrictive limit on $\theta_{13}$ for the global fit. The
allowed 90 and 99\% CL ranges are shown in
Table~\ref{table:limits_comb}.

Adding the solar data to the reactor and atmospheric results also
affects the allowed ranges on the ``non-solar'' parameters
$\theta_{23}$ and $\Delta m^2_{32}$. This effect is illustrated in
panels (b), (c), (e) and (f) of Fig.~\ref{fig:chiglo}.
Figure~\ref{fig:chiglo}.b shows the dependence of
$\Delta\chi^2_{\text{atm + solar}}$ with $\tan^2\theta_{23}$ after
minimizing in the other four parameters.  This figure illustrates that
the allowed $\tan^2\theta_{23}$ range is sensitive to the solar
solution (LMA or SMA) only far in the second octant
($\tan^2\theta_{23}>2$). As discussed in Sec.~\ref{analysis:atmos}
such region of $\theta_{23}$ is favoured for larger values of
$\theta_{13}$. For such $\theta_{13}$ values
$\Delta\chi^2_{\odot,\text{global}}$ depends on the solar solution.
Hence the behaviour observed. As seen in Fig.~\ref{fig:chiglo}.e once
we include the CHOOZ results, the sensitivity on the solar solution
disappears because now we only have very small $\theta_{13}$ values.
Comparing Fig.~\ref{fig:chiglo}.b with Fig.~\ref{fig:chiglo}.e one
sees that the inclusion of CHOOZ data results in a tighter constraint
on $\theta_{23}$ mainly in the second octant irrespective of the
allowed region for the solar parameters.
On the other hand comparing Fig.~\ref{fig:chiglo}.c
with~\ref{fig:chiglo}.f one sees that the lower limit on $\Delta
m^2_{32}$ is rather insensitive to the type of solar neutrino solution
while the upper limit gets reduced and also becomes independent on the
solar solution.

Let us now discuss how the allowed range of solar parameters, $\Delta
m^2_{21}$ and $\theta_{12}$, is affected by the inclusion of the
atmospheric and reactor data in the analysis. In order to illustrate
this point we display in Fig.~\ref{fig:glosol} the allowed regions in
$\Delta m^2_{21}$ and $\theta_{12}$ at 90 and 99\% CL (for 5 degrees
of freedom) after minimizing the function $\chi^2_{\text{atm +
    solar +CHOOZ}}$ with respect to the other three parameters. We
show the figure with two different statistical criteria.

In Fig.~\ref{fig:glosol}.a we show the allowed regions for the
solar  parameters for the unconstrained analysis. The regions are
all defined with respect to the global 5-dimensional minimum (which
occurs in the LMA region) given in Table~\ref{table:chicomb}. This is
the same criterion used to define the regions of the 3-dimensional
solar analysis in Figs.~\ref{fig:3sol_r}--\ref{fig:3sol_rspdn}.
However here, instead of showing the section for a fixed value of
$\theta_{13}$ this parameter has also been minimized in order to
obtain the maximum allowed regions in the 
$(\Delta m^2_{21},\tan\theta_{12})$ plane. As seen from the
figure the allowed parameter space still consists of three well
defined disjoint regions.  This is due to the smallness of
$\theta_{13}$. 
The main difference with the first panel in Fig.~\ref{fig:3sol_rspdn}
arises from the fact that the allowed value of
$\Delta \chi^2$ at a given CL used in the definition of the regions is
now different due to the additional degrees of freedom. 
Also, we see that the LMA region is cut for 
$\Delta m^2_{21}\gtrsim 7\times 10^{-4}$ as a consequence of the inclusion
of the CHOOZ data (see Eq.(\ref{pchooz})).   
 
In Fig.~\ref{fig:glosol}.b each of the three regions: LMA, SMA and LOW
are defined with respect to their local minimum, also given in
Table~\ref{table:chicomb}. These are the parameter ranges in ($\Delta
m^2_{21}, \tan^2\theta_{12}$) allowed by the combined analysis, but
constraining the solar analysis to a given sector.  Notice also that
since the global minimum is in LMA, this region is the same in
Fig.~\ref{fig:glosol}.a and Fig.~\ref{fig:glosol}.b, while the SMA and
LOW regions are different.  

Comparing for example the first panel of Fig.~\ref{fig:3sol_rspdn}
with Fig.~\ref{fig:glosol}.a one sees how the exact size of the
allowed solar regions at a given CL depends on the presence of
additional degrees of freedom. This again illustrates how, as
discussed in Sec.~\ref{analysis:solar}, we cannot infer the actual
quality of the description in a region from its size only.

Summarizing, our final results from the joint solar, atmospheric, and
reactor neutrino data analysis lead to the following allowed ranges of
parameters at 90 [99]~\% CL
\begin{eqnarray}
[1.1\times 10^{-3}]\, 1.4 \times 10^{-3}\;
<& \Delta m^2_{32}/\mbox{\rm eV$^2$}&< \; 
\;6.1\times 10^{-3}\, [7.3\times 10^{-3}] \\ \nonumber
[0.33]\, 0.39<&\tan^2\theta_{23}&<\; 3.1 \, [3.8] \\\nonumber
&\tan^2\theta_{13} &<\;  0.055 \,[0.085] \;\; \text{(Unconstrained or LMA)} \\\nonumber 
&\tan^2\theta_{13} &<\;  0.075 \,[0.135]\;\;  \text{(SMA)} 
\label{globalranges}
\end{eqnarray} 
On the other hand the allowed values of $\Delta m^2_{21}$ and
$\theta_{12}$ can be inferred from Fig.~\ref{fig:glosol}.a or
Fig.~\ref{fig:glosol}.b for unconstrained and constrained cases
respectively.

\section{Summary and Conclusions}
\label{sum}

In this article we have performed a three--flavour analysis of the
full atmospheric, solar and reactor neutrino data. 
The analysis contains five parameters: two mass differences, $\Delta
m^2_{32}$ and $\Delta m^2_{21}$, and three mixing, $\theta_{12}$,
$\theta_{23}$, and $\theta_{13}$. 
Under the assumption of mass hierarchy in neutrino masses
the solar neutrino observables depend
on three of these parameters, which we chose to be $\Delta m^2_{21}$,
$\theta_{12}$ and $\theta_{13}$ while the atmospheric neutrino event
rates depend on $\Delta m^2_{32}$, $\theta_{23}$ and $\theta_{13}$.
The survival probability for reactor neutrinos depends in general 
on four parameters 
$\theta_{12}$, $\Delta m^2_{21}$, $\theta_{13}$, and $\Delta
m^2_{32}$, but for $\Delta m^2_{21}\lesssim 3\times 10^{-4}$ eV$^2$ 
it effectively depends only on
$\theta_{13}$ and $\Delta m^2_{32}$. Thus we have that in 
the hierarchical approximation  the only
parameter common to the three data sets is $\theta_{13}$.

First we have performed independent analyses of the solar neutrino
data and of the atmospheric (and atmospheric plus reactor) neutrino
data in the respective 3--dimensional parameter spaces.  In our solar
data analysis we have studied the dependence of the solutions to the
solar neutrino problem on the $\theta_{13}$ parameter.  We find that
the most favourable scenario for solar neutrino oscillations is the
simplest two--neutrino mixing case, $\theta_{13}=0$ and that for
large enough $\theta_{13}$ angles there is no allowed solution to the
solar neutrino problem. As a result we derive an upper bound on
$\tan^2\theta_{13}< 2.4$ at 90\% CL.

Our results for the three--dimensional analysis of the full
atmospheric neutrino data in the framework of three--neutrino
oscillations show that the most favourable scenario is the $\nu_\mu
\to \nu_\tau$ oscillation hypothesis with the best fit corresponding
to a very small value of $\theta_{13}= 9^\circ$. However a
non-neglegeable $\nu_\mu \to \nu_e$ component is still allowed. The
corresponding upper limit on $\theta_{13}$ is $\tan^2\theta_{13}<
0.34$ at 90\% CL.  From this study we have also derived the allowed
$\Delta m^2_{32}$ and $\tan^2\theta_{32}$ ranges in the framework of
three neutrino mixing.

We have also studied the effect of combining the CHOOZ and the
atmospheric neutrino data in a common three--parameter analysis.  Our
results show that the main effect of the addition of the CHOOZ data to
the atmospheric neutrino analysis is to strengthen the constraint on
the allowed value of the angle $\theta_{13}$ which leads to the upper
limit $\tan^2\theta_{13}< 0.043$ at 90\% CL. This is due to the fact
that the full allowed region of $\Delta m^2_{32}$ from the atmospheric
neutrino analysis lies in the sensitivity range for the CHOOZ reactor.
Including the CHOOZ results has very little effect on the allowed
range of $\Delta m^2_{32}$, while it results into a tighter constrain
on $\tan^2\theta_{23}$ on the second octant.

Finally we have performed the combined five--dimensional
analysis of the solar and atmospheric neutrino data and also in
combination with the reactor results and we have derived the allowed
range of the five parameters. We have also studied how these ranges
depend on the particular solution region for the solar neutrino
deficit. 

In conclusion we see that from our statistical analysis of the solar
data it emerges that the status of the large mixing--type solutions
has been further improved with respect to the previous Super--Kamiokande
data sample, due mainly to the substantially flatter recoil electron
energy spectrum. In contrast, there has been no fundamental change,
other than further improvement due to statistics, on the status of the
atmospheric data.  For the latter the oscillation picture clearly
favours large mixing, while for the solar case the preference is still
not overwhelming. Both solar and atmospheric data favour small
values of the additional $\theta_{13}$ mixing and this behaviour is
strengthened by the inclusion of the reactor limit. 
Nevertheless the prospects that both solar and
atmospheric data select large lepton mixing seems in puzzling
contrast with the observed structure of quark mixing.

\acknowledgments We thank S. Petcov and A. de Gouvea for comments 
and discussions. This work was supported by DGICYT under grants
PB98-0693 and PB97-1261, by Generalitat Valenciana under grant
GV99-3-1-01, and by the TMR network grants ERBFMRXCT960090 and
HPRN-CT-2000-00148 of the European Union.


\begin{table} 
\begin{tabular}{|l|l|l|l|l|} 
Experiment & Rate & Ref. & Units& $ R^{\text BP98}_i $\\ 
\hline 
Homestake  & $2.56\pm 0.23 $ & \protect\cite{chlorine} & SNU &  $7.8\pm 1.1 $   \\ 
GALLEX + SAGE+ GNO  & $74.7\pm 5.2 $ & \protect\cite{gallex,sage} & SNU & $130\pm 7 $  \\ 
Kamiokande & $2.80\pm 0.38$ & \protect\cite{kamioka} &  
$10^{6}$~cm$^-2$~s$^{-1}$ & $5.2\pm 0.9 $ \\    
Super--Kamiokande & $2.40\pm 0.08$ & \protect\cite{sksol00} &  
$10^{6}$~cm$^{-2}$~s$^{-1}$ & $5.2\pm 0.9 $ \\    
\end{tabular} 
\vskip .3cm
\caption{Measured rates for solar neutrinos at 
by Chlorine, Gallium, Kamiokande and Super--Kamiokande experiments. } 
\label{rates} 
\end{table} 

\begin{table}
    \begin{tabular}{|l|c|ccc|cc|}
        Data sets & d.o.f.
        & $\tan^2 \theta_{13}$ & $\tan^2 \theta_{12}$ 
        & $\Delta m^2_{21}~[\text{eV}^2]$
        & $\chi_{\text{SM}}^2$ & $\chi_{\text{min}}^2$ \\
        \hline
    Rates    &$4-3$ &$0.07$ &$1.0\times 10^{-3}$ &$8.2\times 10^{-6}$ 
&$62.2$ &$0.62$  \\
    Spec$_{\text DN}$&$35-3$ &$0.0$ &$0.17$($5.9$) &$4.6\times 10^{-10}$ 
&$30.4$ &$28.0$  \\
    Global   &$39-3$ &$0.0$ &$0.37$ &$3.3\times 10^{-5}$ &$92.6$ &$35.2$  \\
    \end{tabular}
\vskip .3cm
    \caption{Minimum $\chi^2$ values and best-fit points for various sets of 
solar neutrino data.} 
\label{table:chisol}
\end{table}
\begin{table}
    \begin{tabular}{|l|c|ccc|cc|}
        Data sets & d.o.f.
        & $\tan^2 \theta_{13}$ & $\tan^2 \theta_{23}$ 
        & $\Delta m^2_{32}~[\text{eV}^2]$
        & $\chi_{\text{SM}}^2$ & $\chi_{\text{min}}^2$ \\
        \hline
        FINKS            & $10-3$ & 1. & $>100$ & $0.9 \times 10^{-3}$ & $29.9$ & $14.8$ \\
      CONT--UNBIN & $16-3$ & 0.33  & $>100$ & $4.2\times 10^{-3}$ & 
 $58.1$ & $18.4$ \\
      CONT--BIN & $40-3$ & 0.03  &  \01.6 & $2.5 \times 10^{-3}$ &  $115.1$ & $41.3$ \\
      SK CONT
   & $20-3$ & $0.03$  & $0.89$ & $2.3 \times 10^{-3}$ & $80.7$ & $14.5$ \\
      UP--$\mu$   & $25-3$ & 0.23 &  \03.3 & $7.1 \times 10^{-3}$ & $50.3$ & $25.1$ \\
      SK & $35-3$ & 0.005  & 1.3& $2.8 \times 10^{-3}$ &  $131.4$ & $27.8$ \\
        ALL--ATM  & $65-3$ & 0.03  &  \01.6 & $3.3 \times 10^{-3}$ &  $191.7$ & $61.7$ \\ \hline
        ALL--ATM + CHOOZ      & $66-3$ & $0.005$  
&  \01.4 & $3.1 \times 10^{-3}$ &  $191.8$ & $62.5$ \\
    \end{tabular}
    \caption{
Minimum $\chi^2$ values and best-fit points for various sets of 
atmosperic neutrino data}. 
\label{table:chiatm}
\end{table}
\begin{table}
    \begin{tabular}{|l|c|c|c|c|}
        &  & \multicolumn{2}{c|}{$\tan^2\theta_{13}$}  &  $\theta_{13}$ deg \\
        \hline
        & data set           & min   & limit 90\% (99\%) & limit 90\% (99\%)  
\\
        \hline
        Solar 
        & Rates              & 0.22  & 0.8 (2.5)  & $42^\circ$ ($58^\circ$) 
\\
        & Global             & 0.0   & 2.4  (3.5)  & $57^\circ$ ($62^\circ$)   
\\
        \hline
        & FINKS              & 1. & 16.4 (42.7) & $76^\circ$ ($81^\circ$) \\
        & CONT--UNBIN & 0.33  & 10.0 (13.6) & $72^\circ$ ($75^\circ$) \\
        & CONT--BIN      & 0.03  & 0.97 (2.34) & $45^\circ$ ($57^\circ$) \\
        & UP--$\mu$       & 0.23  & 0.55 (7.71) & $37^\circ$ ($70^\circ$) \\
        & SK  & 0.005 & 0.33 (0.68) & $30^\circ$ ($40^\circ$) \\
        & ALL--ATM                & 0.03 & 0.34 (0.57) & $30^\circ$ 
($37^\circ$) \\
        \hline
        \multicolumn{2}{|c|} {ALL--ATM + Chooz} 
        & 0.005   & 0.043 (0.08) & $12^\circ$ ($16^\circ$) \\
    \end{tabular}
    \vskip .3cm
\caption{Upper bounds on $\theta_{13}$ at 90 and 99\% CL from the analysis of
the different smaples and combinations of atmosperic neutrino data.}
\label{table:limits}
\end{table}
\begin{table}
    \begin{tabular}{|l|c|ccccc|c|}
data set&  Solar region  
        & $\tan^2 \theta_{13}$ 
        & $\tan^2 \theta_{12}$ 
        & $\Delta m^2_{21}~[\text{eV}^2]$ 
        & $\tan^2 \theta_{23}$ 
        & $\Delta m^2_{32}~[\text{eV}^2]$
        & $\chi_{\text{min}}^2$ \\
        \hline
Solar + ALL--ATM.& Unconst=LMA
        & $0.015$  & $0.35$ & $3.3\times 10^{-5}$ & $1.6$ & $3.3\times 10^{-3}$& $97.1$\\
  & LOW
        & $0.025$  & $0.70$ & $9.2\times 10^{-8}$ & $1.6$ & $3.3\times 10^{-3}$& $100.8$\\
  & SMA
        & $0.03$  & $5.8\times 10^{-4}$ & $5.6\times 10^{-6}$ & $1.7$ & $3.3\times 10^{-3}$& $101.5$\\\hline
Solar + ALL--ATM.& Unconst=LMA
        & $0.005$  & $0.36$ & $3.3\times 10^{-5}$ & $1.4$ & $3.1\times 10^{-3}$& $97.7$\\
+CHOOZ 
  & LOW
        & $0.005$  & $0.58$ & $9.6\times 10^{-8}$ & $1.4$ & $3.1\times 10^{-3}$& $101.4$\\
  & SMA
        & $0.005$  & $6.8\times 10^{-4}$ & $5.1\times 10^{-6}$ & $1.4$ & $3.1\times 10^{-3}$& $103.1$\\
    \end{tabular}
    \vskip .3cm
\caption{Minimum $\chi^2$ values and best-fit points for various sets 
the combined analysis of atmospheric, solar and reactor neutrino data.}
\label{table:chicomb}
\end{table}
\begin{table}
    \begin{tabular}{|l|c|c|c|c|}
  \multicolumn{2}{|c|}{ }      & \multicolumn{3}{|c|}{limit 90\% (99\%)}\\
        \hline
data set&  Solar region & $\tan^2\theta_{13}$  &  $\tan^2\theta_{23}$ 
&  $\Delta m^2_{32}~[\text{eV}^2]$ \\
        \hline
           & Unconstrained
        & $<0.32 (<0.52)$   & $[0.39,4.3]([0.33,6.5])$ & 
$[1.35,6.5]([1.1,8.0])\times 10^{-3}$\\
Solar + ALL--ATM.& LMA
        & $<0.20 (<0.39)$   & $[0.39,3.6]([0.33,5.2])$  &
$[1.35,6.5]([1.07,8.0])\times 10^{-3}$\\
  & SMA
        & $<0.46 (<0.75)$   & $[0.40,6.0]([0.34,8.3])$  &
$[1.35,7.5]([1.1,9.7])\times 10^{-3}$\\
\hline Solar + ALL--ATM & Unconst=LMA
& $<0.055 (<0.085)$& $[0.39,3.0]([0.33,3.7])$ & 
$[1.4,6.1]([1.1,7.3])\times 10^{-3}$\\
+ CHOOZ  & SMA 
& $<0.075 (<0.135)$& $[0.39,3.1]([0.33,3.8])$ & 
$[1.3,6.1]([1.0,7.3])\times 10^{-3}$\\
    \end{tabular}
    \vskip .3cm
\caption{Allowed ranges on $\tan^2\theta_{13}$, $\tan^2\theta_{23}$ and 
$\Delta m^2_{32}~[\text{eV}^2]$ at 90 and 99\% CL from the
combined analysis.}
\label{table:limits_comb}
\end{table}
%

%
\newpage

\end{document}